\documentclass[11pt]{article}

\usepackage[final]{acl}

\usepackage{times}
\usepackage{latexsym}

\usepackage[T1]{fontenc}
\usepackage[utf8]{inputenc}

\usepackage{microtype}

\usepackage{inconsolata}

\usepackage{graphicx}
\usepackage{amsmath,graphicx,hyperref,algorithm,algpseudocode,placeins,booktabs,multirow,amsmath,xcolor,amssymb}
\title{SEMAG: Self-Evolutionary Multi-Agent Code Generation}

\author{Yulin Peng \textsuperscript{\rm 1}, Haowen Hou\textsuperscript{\rm 2}, Xinxin Zhu \textsuperscript{\rm 1, \rm 2}, \\{\bf Ying Tiffany He} \textsuperscript{\rm 1}{\bf, F. Richard Yu} \textsuperscript{\rm 3} \\
  \textsuperscript{\rm 1}College of Computer Science and Software Engineering, Shenzhen University, China \\
  \textsuperscript{\rm 2}Guangdong Laboratory of Artificial Intelligence and Digital Economy (SZ), China \\
  \textsuperscript{\rm 3}School of Information Technology, Carleton University, Canada \\
  }

\begin{document}
\maketitle
\begin{abstract}
Large Language Models (LLMs) have made significant progress in handling complex programming tasks. However, current methods rely on manual model selection and fixed workflows, which limit their ability to adapt to changing task complexities. To address this, we propose SEMAG, a Self-Evolutionary Multi-Agent code Generation framework that mimics human coding practices. It decomposes programming tasks into stages, including planning, coding, debugging, and discussion, while adapting workflows to task difficulty. Its self-evolutionary agents can access the latest models in real time and automatically upgrade the backbone model. SEMAG sets new state-of-the-art Pass@1 accuracy across benchmarks. Using identical backbone models, SEMAG outperforms prior methods by 3.3\% on CodeContests. When augmented with self-evolutionary model selection that automatically identifies optimal backbones, SEMAG reaches 52.6\%, showcasing both framework effectiveness and adaptability to evolving LLM capabilities.
\end{abstract}

\section{Introduction}

Large Language Models (LLMs) have demonstrated substantial progress in code generation and completion, driven by large-scale pretraining on diverse codebases. The GPT series \cite{achiam2023gpt, hurst2024gpt}, CodeLLaMA-2 \cite{roziere2023code}, Qwen2.5-Coder \cite{hui2024qwen2}, and DeepSeek-v3 \cite{liu2024deepseek} exhibit strong coding capabilities,  unlocking new avenues for automated software development. In parallel, multi-agent frameworks and debugging-enhanced methodologies—such as planning-centric workflows \cite{lei2024planning}, self-debugging paradigms \cite{chen2023teaching}, and collaborative agent systems \cite{zhong-etal-2024-debug}—have shown promising performance on standard benchmarks. Nonetheless, real-world scenarios present open-ended tasks, constrained computational budgets, and evolving specifications, revealing critical limitations in current approaches.

First, frameworks such as Self-Debugging \cite{chen2023teaching}, LDB \cite{zhong-etal-2024-debug} typically adopt a \textit{fixed reasoning depth}. On simple tasks, they introduce unnecessarily complex workflows, leading to redundant computation and excessive token usage, while on difficult tasks, the shallow reasoning depth results in poor success rates. Although hierarchical prompting has been shown to mitigate unnecessary reasoning \cite{budagam2024hierarchical}, these approaches still lack a principled mechanism to adapt reasoning depth dynamically to task complexity.  

\begin{figure}[!t]
    \centering
    \includegraphics[width=\linewidth]{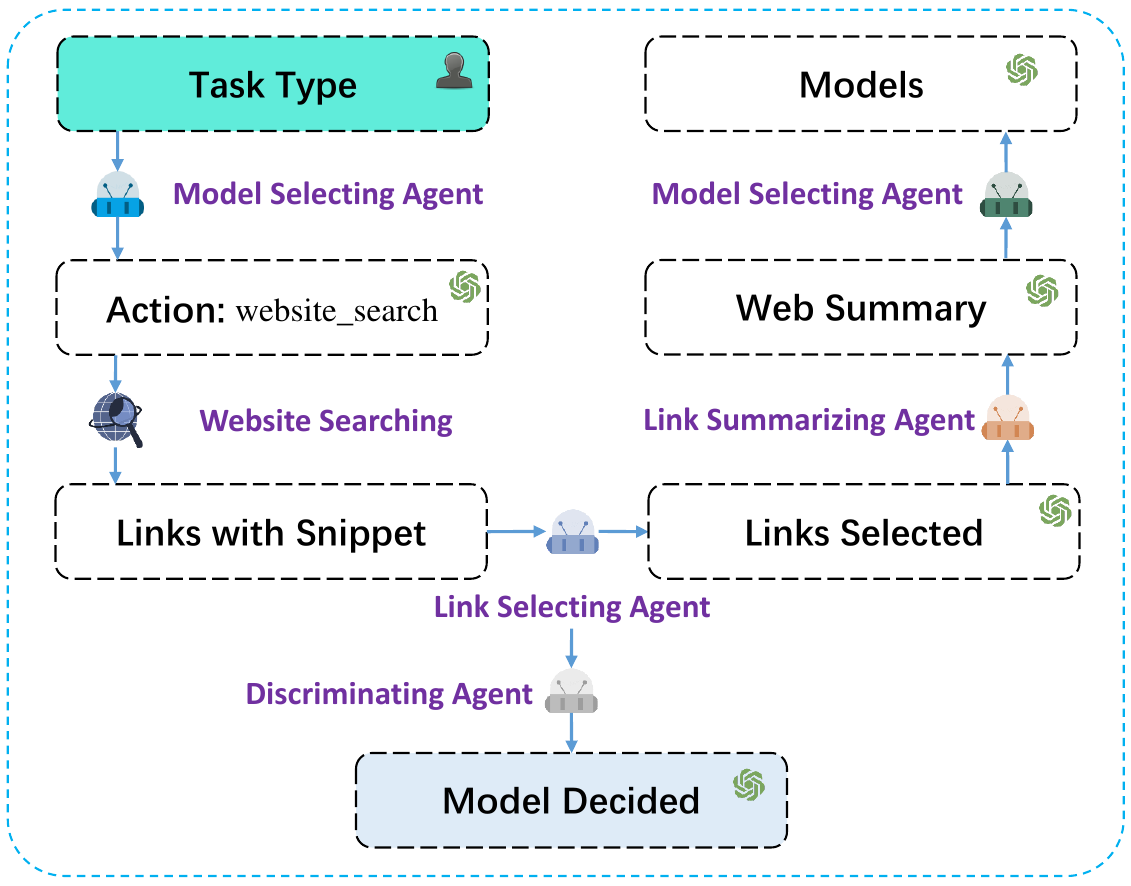}
    \caption{Overview workflow of Self-Evolution Agents. Agents integrate insights from recent research, news, and community discussions, dynamically identify and deploy the most suitable models.}
    \label{fig:overview_self-evolution}
\end{figure}
\begin{figure*}[t]
    \centering
    \includegraphics[width=\textwidth]{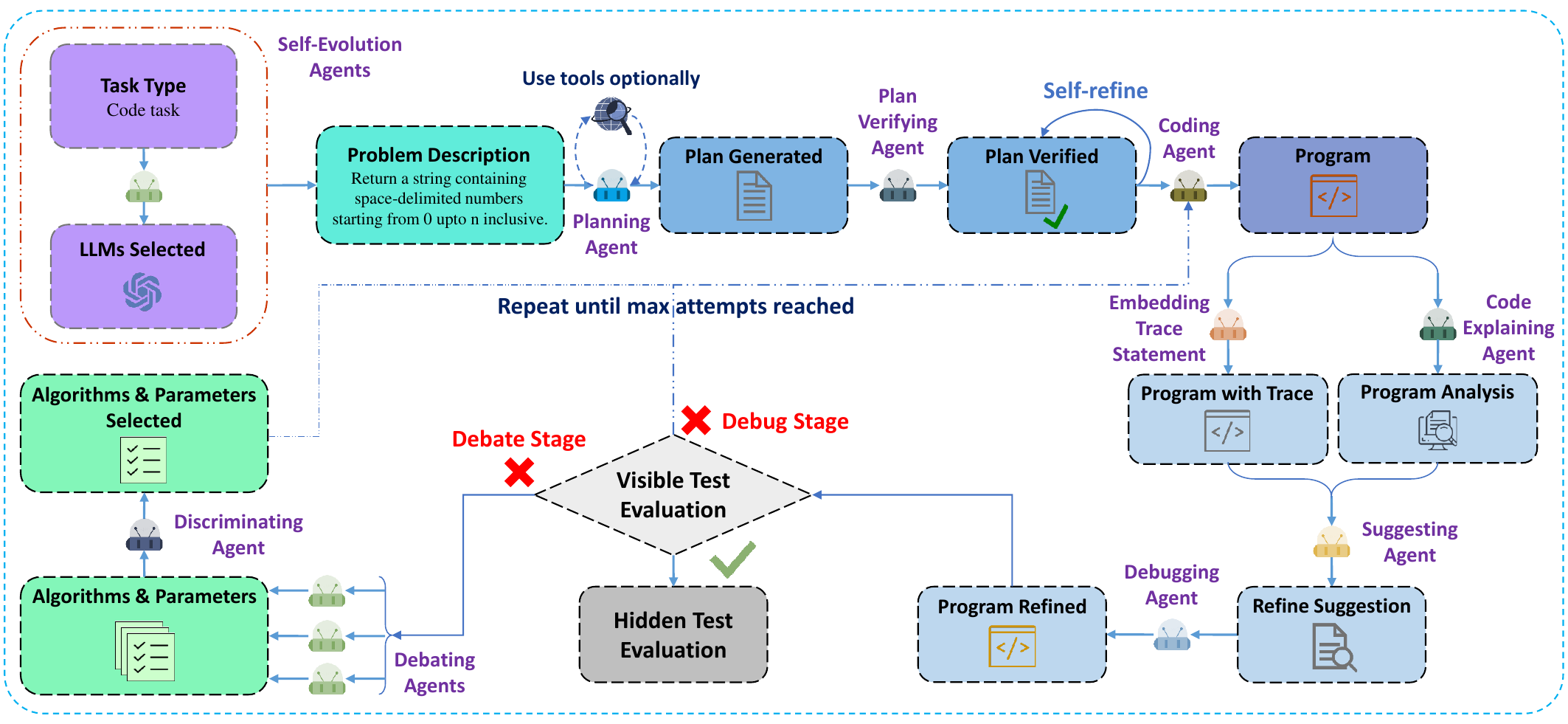}
    \caption{Overview of SEMAG. (1) \textbf{Self-Evolve:} Agents dynamically select optimal backbone LLMs per task requirements. (2) \textbf{Plan:} Planning Agent creates solution plans validated by Plan Verifying Agent through I/O simulation. (3) \textbf{Debug:} Coding Agent generates code; upon failure, specialized agents (Embedding Trace, Code Explaining, Suggesting, Debugging) collaboratively refine using trace logs. (4) \textbf{Debate:} When debugging stalls, Debating Agents propose alternatives with Discriminating Agent selecting the optimal configuration.}
    \label{fig:overview_code_agent}
\end{figure*}
Second, current pipelines utilize a single debugging iteration. When initial outputs diverge significantly from the target, systems are prone to local minima. Though advanced reasoning paradigms such as Chain-of-Thought \cite{wei2022chain}, Tree-of-Thoughts \cite{yao2023tree}, and parallel candidate exploration \cite{li2025s} enhance complex reasoning, they lack explicit discussion–decision phases that aggregate diverse reasoning trajectories for improved synthesis.

Third, most systems are tightly coupled to a single backbone model. Frameworks built on GPT \cite{achiam2023gpt, hurst2024gpt}, Gemini \cite{team2023gemini, team2024gemini}, or Claude \cite{claude} typically depend on a static model throughout execution. As task characteristics shift or new models emerge, backbone switching often requires manual intervention, limiting adaptability and scalability.

To address these challenges, we propose \textbf{SEMAG}, a \textbf{S}elf-\textbf{E}volutionary \textbf{M}ulti-\textbf{A}gent code \textbf{G}eneration framework. Our contributions are summarized as follows: 
\begin{itemize}
\item \textbf{Adaptive hierarchical prompting:} We propose a dynamic strategy that adjusts reasoning depth based on task complexity.
\item \textbf{Collaborative self-evolution:} We introduce discussion–decision module enabling escape from local optima and adaptive backbone switching. 
\item \textbf{Empirical gains:} Achieves state-of-the-art performance on seven benchmarks. With controlled backbone comparison, SEMAG improves 3.3\% over the previous best method on CodeContests; with self-evolutionary model selection, it further reaches 52.6\%.
\end{itemize}

We evaluate SEMAG across seven text-to-code benchmarks, including four foundational datasets (HumanEval, MBPP, HumanEval-ET, MBPP-ET) and three competition-level benchmarks (APPS, LiveCode, CodeContests). Experimental results show that SEMAG achieves new state-of-the-art performance, including 98.8\% Pass@1\cite{chen2021evaluating,dong2024self} on HumanEval, 87.6\% on MBPP, and 65.0\% on LiveCode. Most notably, on the most challenging dataset CodeContests, SEMAG achieves 38.0\% Pass@1 accuracy with GPT-4o (3.3\% improvement over LPW under the same backbone). When augmented with self-evolutionary model selection that automatically identifies the optimal backbone, SEMAG further reaches 52.6\%. These results demonstrate that SEMAG achieves superior performance and resource efficiency, while offering strong adaptability to evolving programming tasks.

\section{Related Work}

\subsection{Traditional Approaches to Program Synthesis}

Program synthesis has a long-standing research foundation in artificial intelligence \cite{waldinger69, manna1971toward}. Traditional methods leverage search strategies and data flow analysis \cite{mccarthy1978history}. Early efforts aimed to advance automatic programming and to identify viable approaches \cite{balzer1985, soloway1986} or explore large program spaces through domain-specific languages \cite{mernik2005and, gu2021domain}. These approaches struggle with generalization and scalability due to search space complexity.

\subsection{Large Language Models for Code Synthesis}

Pretrained language models have enhanced code synthesis, with specialized models such as Qwen2.5-Coder \cite{hui2024qwen2}, CodeLLaMA-2 \cite{roziere2023code}, Mistral \cite{jiang2024mixtral}, and DeepSeek-v3 \cite{liu2024deepseek} excelling in programming tasks. General-purpose models, including GPT \cite{achiam2023gpt, hurst2024gpt}, Gemini \cite{team2023gemini, team2024gemini}, and Claude \cite{claude}, also demonstrate robust code generation capabilities. However, these models still face challenges related to syntactic correctness, semantic alignment, generation robustness, and version conflicts. As a result, more refined control and evaluation mechanisms for code generation are necessary.

\subsection{Prompting and Debugging Techniques}

Researchers have proposed various prompting and debugging techniques to improve code generation. Prompting strategies generally fall into three categories: retrieval-based \cite{islam2024mapcoder}, planning-based \cite{yao2022react}, and debugging-based \cite{chen2023teaching} approaches. These aim to guide LLMs in decomposing complex tasks into manageable parts through step-by-step reasoning. Techniques such as Chain-of-Thought \cite{wei2022chain}, Tree-of-Thoughts \cite{yao2023tree}, and cumulative reasoning mimic human problem-solving paths, significantly enhancing model performance on complex tasks \cite{zhou2022least,zhang2023cumulative}. More advanced methods simulate the software development process by constructing multiple candidate programs and exploring the solution space in parallel \cite{li2025s,antoniades2024swe}.

Debugging systems such as Self-Debugging \cite{chen2023teaching} and LDB \cite{zhong-etal-2024-debug} iteratively refine code using model explanations, execution, and human feedback. However, their effectiveness decreases when the initial code diverges from the intended function. To improve generation quality with limited supervision, some methods break down the coding task by incorporating visible test cases, step-by-step verification \cite {hu2025qualityflow,li2024large,mathews2024test}, and natural language instructions to improve controllability and alignment.

Previous methods either fix reasoning depth—wasting compute on simple tasks and underperforming on hard ones—or rely on a single LLM, limiting recovery from failures. SEMAG tackles both with three mechanisms: a hierarchical controller that scales from one-shot to multi-step planning based on feedback; a discussion–decision phase where agents critique and merge solutions to avoid local optima; and an automatic model selector that switches to a more capable backbone, boosting Pass@1 accuracy as difficulty rises.

\section{Method}

We present a hierarchical multi-agent framework for code synthesis that adapts to task complexity through progressive refinement levels, coupled with a self-evolution mechanism for dynamic model selection. The overview of SEMAG is shown in Figure \ref{fig:overview_code_agent}.

\subsection{Problem Formulation}
We define a code generation task as $\mathcal{T} = (P, S, \mathcal{C})$ where $P \in \mathcal{P}$ is problem description, $S = \{(x_i, y_i)\}_{i=1}^n$ are input-output examples, and $\mathcal{C}$ is the program space. The core agent operations are:
{

\begin{equation}
\begin{split}
\text{CODER} &: \mathcal{P} \times \mathcal{S} \times \Pi \times \Theta \rightarrow \mathcal{C}, \\ \text{PLANNER} &: \mathcal{P} \times \mathcal{S} \rightarrow \Pi, \\ \text{VERIFIER}&: \Pi \times \mathcal{P} \times \mathcal{S} \rightarrow {0,1} \times \Pi \times \mathcal{L}, \\ \text{DEBUGGER}&: \mathcal{C} \times \Sigma \rightarrow \mathcal{C}
\end{split}
\end{equation}
}

where $\Pi$ is the plan space, $\Theta$ parameters, $\mathcal{L}$ logs, and $\Sigma$ suggestions. Additional agents include EMBEDTRACE ($\mathcal{C} \rightarrow \mathcal{T}$), EXPLAINER ($\mathcal{C} \times \mathcal{P} \rightarrow \mathcal{E}$), and SUGGESTOR ($\mathcal{T} \times \mathcal{L} \times \mathcal{E} \rightarrow \Sigma$).

\subsection{Hierarchical Code Synthesis Framework}
Our framework employs a four-level hierarchical architecture that progressively increases computational effort based on task complexity. 

\noindent \textbf{Level 1 (Direct Generation):} The system initially attempts direct code synthesis using minimal prompting:
{
\begin{equation}
Y = \text{CODER}(P, S, \varnothing, \varnothing),
\end{equation}
}

where $\varnothing$ indicates no plan or parameters. 

\noindent \textbf{Level 2 (Planning and Verification):} Upon Level 1 failure, the system generates and iteratively refines a structured solution plan. The planning process operates as:
{
\begin{equation}
\pi_0 = \text{PLANNER}(P, S),
\end{equation}
}

followed by iterative verification:
{\small
\begin{equation}
\begin{split}
(\nu_i, \pi_i, \ell_i) = \text{VERIFIER}(\pi_{i-1}, P, S), \\
& i \in [1, M_{\text{plan}}].
\end{split}
\end{equation}
}

where $\nu_i \in \{0,1\}$ indicates verification status, $\pi_i$ is the refined plan, and $\ell_i$ contains verification logs. The process terminates when $\nu_i = 1$ or $i = M_{\text{plan}}$, with the final plan $\pi^*$ guiding code generation:
{
\begin{equation}
Y = \text{CODER}(P, S, \pi^*, \varnothing).
\end{equation}
}
\begin{algorithm}[t]
\small
\caption{Hierarchical workflow of SEMAG}
\label{algo:semag}
\textbf{Input}: Problem $P$, examples $S$ \\
\textbf{Output}: Program $Y$
\begin{algorithmic}[1]

\State $Y \gets \text{CODER}(P,S)$\Comment\textbf{Level 1}
\If{$\textsc{Test}(Y,S)$} \Return $Y$ \EndIf

\State $\pi \gets \text{PLANNER}(P,S)$ \Comment\textbf{Level 2}
\For{$i=1$ to $M_{\text{plan}}$}
    \State $(\nu,\pi,\ell) \gets \text{VERIFIER}(\pi,P,S)$
    \If{$\nu=1$} \textbf{break} \EndIf
\EndFor
\State $Y \gets \text{CODER}(P,S,\pi)$
\If{$\textsc{Test}(Y,S)$} \Return $Y$ \EndIf

\For{$t=1$ to $M_{\text{try}}$} \Comment\textbf{Level 3}
    \State $\tau_{\text{prev}} \gets \varnothing$
    \For{$d=1$ to $M_{\text{debug}}$}
        \State $\tau \gets \text{EMBEDTRACE}(Y)$
        \State $\sigma \gets \text{SUGGESTOR}(\tau,\ell,\text{EXPLAINER}(Y,P))$
        \State $Y \gets \text{DEBUGGER}(Y,\sigma)$
        \If{$\textsc{Test}(Y,S)$} \Return $Y$ \EndIf
        \If{$\rho(\tau,\tau_{\text{prev}}) > \delta(d,\mathcal{T})$} \textbf{break} \EndIf
        \State $\tau_{\text{prev}} \gets \tau$
    \EndFor

    \State $H \gets \{\text{DEBATER}_j(P,\tau,Y)\}_{j=1}^{N_{\text{debater}}}$\ \Comment\textbf{Level 4}
    \State $Y \gets \text{CODER}(P,S,\text{DECIDER}(H))$
    \If{$\textsc{Test}(Y,S)$} \Return $Y$ \EndIf
\EndFor

\State \Return $Y$

\end{algorithmic}
\end{algorithm}

\noindent \textbf{Level 3 (Trace-Guided Debugging):} When Level 2 fails, the system enters an iterative debugging phase with $K_{\text{pass}}$ passes and $M_{\text{try}}$ attempts per pass. For each attempt, the debugging process consists of:
{
\begin{equation}
\begin{split}
\tau = \text{EMBEDTRACE}(Y),\\ \qquad 
\epsilon = \text{EXPLAINER}(Y, P), \\
\sigma = \text{SUGGESTOR}(\tau, \ell^*, \epsilon),\\ \qquad
Y' = \text{DEBUGGER}(Y, \sigma).
\end{split}
\end{equation}
}
This process repeats for $M_{\text{debug}}$ iterations, where $\tau$ captures runtime variable states, $\epsilon$ provides semantic analysis, and $\sigma$ synthesizes targeted modifications.

\noindent \textbf{Level 4 (Multi-Agent Collaborative Refinement):} When iterative debugging stalls, the system employs collaborative multi-agent discussion. Each of $N_{\text{debater}}$ agents generates proposals incorporating discussion history:
{
\begin{equation}
\begin{split}
d_j = \text{DEBATER}_j(P, \tau, Y, H_{j-1}), \\
j &\in [1, N_{\text{debater}}].
\end{split}
\end{equation}
}
where $H_{j-1} = \{d_1, ..., d_{j-1}\}$ represents accumulated discussion history. The decision aggregation employs weighted consensus:
{
\begin{equation}
\begin{split}
(\alpha^*, \theta^*) 
= \arg\max_{(\alpha, \theta)} 
\sum_{j=1}^{N_{\text{debater}}} 
w_j \cdot \phi(d_j, \alpha, \theta), \\
w_j = \frac{\exp(\eta_j / \tau_w)}
{\sum_k \exp(\eta_k / \tau_w)}.
\end{split}
\end{equation}
}
where $\eta_j$ represents historical performance and $\phi$ evaluates proposal alignment.

\begin{table*}[h]  
\centering  
\small
\setlength{\tabcolsep}{8pt}
\renewcommand{\arraystretch}{1}
\begin{tabular}{cl|cccc}  
\toprule  
Model & Method & HumanEval & MBPP & HumanEval-ET & MBPP-ET \\
\midrule  
\multirow{7}{*}{GPT-3.5}   
 & Direct & 72.0\% $\pm$ 1.2\% & 55.2\% $\pm$ 0.8\% & 62.8\% $\pm$ 0.6\% & 45.6\% $\pm$ 0.6\% \\
 & Self-Planning & 77.4\% $\pm$ 1.8\% & 69.2\% $\pm$ 0.4\% & 69.5\% $\pm$ 0.6\% & 52.4\% $\pm$ 1.0\% \\
 & MapCoder & 77.4\% $\pm$ 0.6\% & 72.0\% $\pm$ 0.6\% & 66.5\% $\pm$ 1.2\% & 56.6\% $\pm$ 0.8\% \\
 & LDB & 81.1\% $\pm$ 0.6\% & 72.4\% $\pm$ 0.2\% & 72.6\% $\pm$ 1.8\% & 55.6\% $\pm$ 0.4\% \\
 & LPW & 89.0\% $\pm$ 0.8\% & 76.0\% $\pm$ 0.2\% & 77.4\% $\pm$ 0.8\% & 57.6\% $\pm$ 0.2\% \\
 &\multirow{2}{*}{\textbf{SEMAG (Ours)}}  & \textbf{91.5\% $\pm$ 1.8\%}  & \textbf{76.2\% $\pm$ 0.8\%}  & \textbf{79.9\% $\pm$ 0.6\%}  & \textbf{64.4\% $\pm$ 0.4\%}  \\
 & & (+27.1\%) & (+38.0\%) & (+27.2\%) & (+41.2\%) \\
\bottomrule  
\end{tabular}
\caption{Pass@1 accuracy comparison of different methods using GPT-3.5 on code generation benchmarks. The values enclosed in parentheses represent the improvement over the Direct Prompting approach. The standard deviation ($\pm$) is calculated based on the results of three independent runs and applies to the data analysis of subsequent experiments.}  
\label{tab:gpt35_results}  
\end{table*}

\subsection{Adaptive Level Transition Mechanism}
Rather than using fixed iteration thresholds, we employ an adaptive transition mechanism based on execution trace similarity. The transition decision is formulated as:
{
\small
\begin{equation}
\text{Transition}(t) = \begin{cases}
\text{True}, \text{if } \rho(\tau_t, \tau_{t-1}) > \delta(t, \mathcal{T}) \\
\text{False}, \text{otherwise}
\end{cases}
\end{equation}
}

where $\rho$ measures trace similarity using normalized edit distance:
{

\begin{equation}
\rho(\tau_t, \tau_{t-1}) = 1 - \frac{\text{EditDist}(\tau_t, \tau_{t-1})}{\max(|\tau_t|, |\tau_{t-1}|)}
\end{equation}
}

The adaptive threshold $\delta(t, \mathcal{T})$ adjusts based on task complexity and iteration count:
{

\begin{equation}
\delta(t, \mathcal{T}) = \delta_0 \cdot \exp\left(-\lambda \cdot \frac{t}{T_{\max}} \cdot \text{complexity}(\mathcal{T})\right)
\end{equation}
}

where $\delta_0 = 0.85$ is the initial threshold, $\lambda = 0.5$ is the decay rate, $t \in [1, T_{\max}]$ is the current iteration count within the active level, and $T_{\max}$ represents the maximum iterations before mandatory level transition.

\subsection{Self-Evolution Mechanism}
To enable dynamic adaptation to evolving LLMs, we propose an automated model selection framework employing $N_{\text{selectors}}$ parallel agents. Each selector $i$ performs four operations:
First, it generates task-specific keywords $\kappa_i = \text{KEYWORDGEN}(T, \text{context})$ and retrieves recent information $L_i = \text{SEARCH}(\kappa_i)$ by searching tools. Then, relevant links are filtered and summarized:
{

\begin{equation}
    \quad L'_i = \{l \in L_i : \text{relevance}(l, T) > \theta_r\},
\end{equation}
}

{
\begin{equation}
    C_i = \bigcup_{\ell \in L'_i} \text{SUMMARIZE}(\ell).
\end{equation}
}

Third, each selector proposes models $m_i$ with confidence score:
{
\small
\begin{equation}
(m_i, r_i, s_i) = \text{SELECTOR}(C_i, \text{Perf}(m_i, T_{\text{sample}})),
\end{equation}
}

where $s_i$ reflects sampled performance on task subset $T_{\text{sample}}$. Finally, consensus is achieved through weighted voting:
\begin{equation}
m^* = \arg\max_{m \in M} \sum_{i=1}^{N_{\text{selectors}}} s_i \cdot \mathbb{I}[m_i = m].
\end{equation}
This mechanism ensures optimal model selection without manual intervention while maintaining adaptability to emerging LLMs.

\begin{table*}[h]  
\centering  
\small

% \renewcommand{\arraystretch}{1}
% ======= Part 1: 前四列 =======
% ======= Part 1 (等宽) =======
\begin{tabular*}{0.85\linewidth}{cl|cccc}
\toprule
Model & Method
    & HumanEval & MBPP 
    & HumanEval-ET
    & MBPP-ET \\
\midrule
\multirow{5}{*}{GPT-4o}
 & Direct 
    & 91.5\% $\pm$ 1.8\% 
    & 62.8\% $\pm$ 0.4\%
    & 79.3\% $\pm$ 1.2\%
    & 51.0\% $\pm$ 0.2\% \\
 & LDB 
    & 92.1\% $\pm$ 1.2\%
    & 82.4\% $\pm$ 0.8\%
    & 81.7\% $\pm$ 1.8\%
    & 65.4\% $\pm$ 1.0\% \\
 & LPW 
    & 98.2\% $\pm$ 0.6\%
    & 84.8\% $\pm$ 0.6\%
    & 84.8\% $\pm$ 1.2\%
    & 65.8\% $\pm$ 0.8\% \\
 &\multirow{2}{*}{\textbf{SEMAG (Ours)}}
    & \textbf{98.8\% $\pm$ 0.6\%}
    & \textbf{87.6\% $\pm$ 0.4\%}
    & \textbf{86.6\% $\pm$ 0.6\%}
    & \textbf{71.8\% $\pm$ 0.2\%} \\
 & 
    & (+8.0\%) & (+38.9\%) & (+9.2\%) & (+40.8\%) \\
\bottomrule
\end{tabular*}

\vspace{0.35cm}

% ======= Part 2 (等宽) =======
\begin{tabular*}{0.85\linewidth}{cl|cccc}
\toprule
Model & Method & APPS & LiveCode & CodeContests & Overall Avg. \\
\midrule
\multirow{5}{*}{GPT-4o}
 & Direct
    & 47.5\% $\pm$ 0.3\%
    & 46.4\% $\pm$ 0.8\%
    & 24.6\% $\pm$ 1.3\%
    & 57.6\% \\
 & LDB
    & 53.2\% $\pm$ 0.7\%
    & 54.3\% $\pm$ 0.7\%
    & 29.3\% $\pm$ 0.7\%
    & 65.5\% \\
 & LPW
    & 62.6\% $\pm$ 0.3\%
    & 59.3\% $\pm$ 1.4\%
    & 34.7\% $\pm$ 0.7\%
    & 70.0\% \\
  &\multirow{2}{*}{\textbf{SEMAG (Ours)}}
    & \textbf{67.6\% $\pm$ 0.8\%}
    & \textbf{65.0\% $\pm$ 0.7\%}
    & \textbf{38.0\% $\pm$ 1.3\%}
    & \textbf{73.6\%} \\
 & 
    & (+42.3\%) & (+40.1\%) & (+54.5\%) & (+27.7\%) \\
\bottomrule
\end{tabular*}

\caption{Pass@1 accuracy comparison of different methods using GPT-4o (2024-05-13) across multiple benchmarks. The values enclosed in parentheses represent the improvement over the Direct Prompting approach.}  
\label{tab:gpt4o_results}  
\end{table*}

\begin{table*}[h] 
\small
\centering  
\renewcommand{\arraystretch}{1}
\begin{tabular}{l|ccccccc}  
\toprule  
\multirow{2}{*}{Level} & \multicolumn{7}{c}{Benchmark} \\
\cmidrule(lr){2-8}  
 & HumanEval & MBPP & HumanEval-ET & MBPP-ET & APPS & LiveCode & CodeContests \\
\midrule  
Level 1 & 148 & 314 & 130 & 255 & 66 & 65 & 37 \\
Level 2 & 8 & 18 & 6 & 10 & 9 & 16 & 6 \\
Level 3 & 4 & 48 & 2 & 46 & 7 & 4 & 5 \\
Level 4 & 4 & 120 & 26 & 189 & 57 & 55 & 102 \\
\bottomrule  
\end{tabular}  
\caption{Distribution of prompt difficulty levels across multiple benchmarks using GPT-4o (2024-05-13).}  
\label{tab:level_distribution}  
\end{table*} 

\section{Experiments}
\subsection{Experimental Setup}

\noindent \textbf{Evaluation Datasets.} We evaluate SEMAG on seven text-to-code benchmarks across two categories. The foundational datasets include HumanEval \cite{chen2021evaluating} and HumanEval-ET (164 problems each), and MBPP \cite{austin2021program} and MBPP-ET (500 problems each). The ET variants \cite{dong2025codescore} extend their counterparts with additional edge test cases. For MBPP/MBPP-ET, which lack sample input-output pairs, we follow previous work \cite{zhong-etal-2024-debug, lei2024planning} by randomly selecting one test case from the hidden test set as a sample (excluded from evaluation). The competition-level datasets consist of APPS \cite{hendrycks2021measuring} (139 problems), LiveCode \cite{jain2024livecodebench} (140 problems), and CodeContests \cite{li2022competition} (150 problems). LiveCode, released after the LLM training cutoff, ensures uncontaminated evaluation.

\noindent \textbf{Baseline Methods.} We compare SEMAG against several baseline approaches: \textit{Direct} inputs tasks directly into an LLM; \textit{Self-Planning} \cite{jiang2023self} decomposes tasks into subgoals; \textit{MapCoder} \cite{islam2024mapcoder} employs four agents for retrieval, planning, execution, and debugging; \textit{LDB} \cite{zhong-etal-2024-debug} utilizes control flow diagrams for programme decomposition and error localization; and \textit{LPW} \cite{lei2024planning}, the state-of-the-art approach, verifies plans step-by-step and uses print statements for debugging.

\subsection{Main Results}

\noindent \textbf{Comparison with Baselines.} Tables \ref{tab:gpt35_results} and \ref{tab:gpt4o_results} present results using GPT-3.5 and GPT-4o as backbone models. With GPT-3.5, SEMAG achieves the highest Pass@1 accuracy across all benchmarks, outperforming the strongest baseline LPW by 2.5\%, 0.2\%, 2.5\%, and 6.8\% on HumanEval, MBPP, HumanEval-ET, and MBPP-ET respectively.

Using GPT-4o, SEMAG establishes new state-of-the-art results across all seven benchmarks, achieving 98.8\% accuracy on HumanEval (solving 162/164 problems). Compared to LPW, SEMAG demonstrates consistent improvements of 1.8-6.0\% on foundational benchmarks and 3.3-5.7\% on competition-level benchmarks, with particularly significant gains of 40-54\% over Direct prompting.

\begin{figure}
    \centering
    \includegraphics[width=0.9\linewidth]{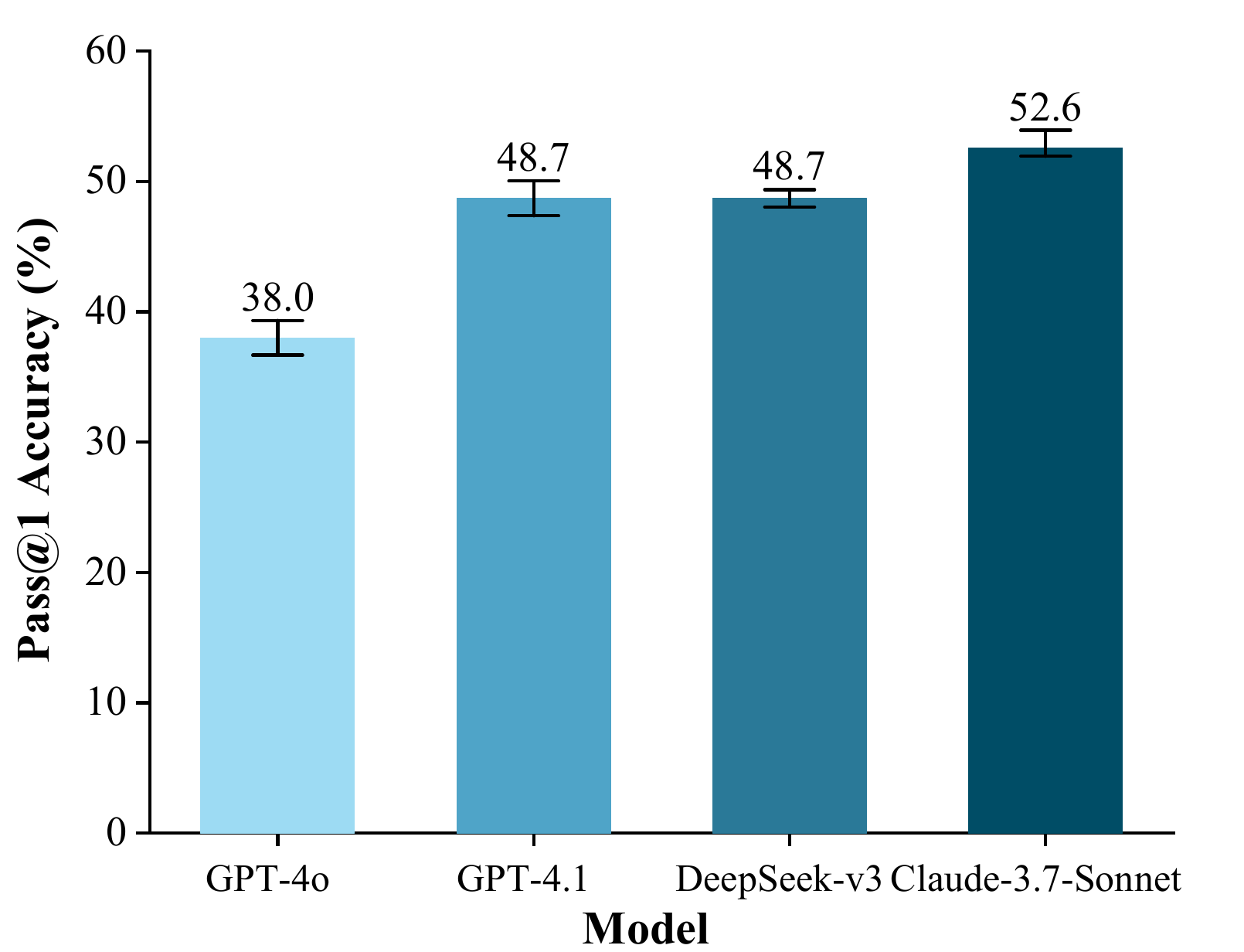}
    \caption{Pass@1 accuracy on CodeContests using GPT-4o(2024-05-13), GPT-4.1(2025-04-14), DeepSeek-v3(2025-03-24) and Claude-3.7-Sonnet(2025-02-19).}
    \label{fig:find_model_result}
\end{figure}

\noindent \textbf{Self-Evolution Agents in Code Task.} To evaluate self-evolution capability, we deploy agents on the CodeContests benchmark to select optimal LLMs autonomously. Agents analyze real-time information to identify three candidate models: Claude-3.7-Sonnet, GPT-4.1, and DeepSeek-v3. Figure \ref{fig:find_model_result} shows that Claude-3.7-Sonnet achieves 52.6\% Pass@1 accuracy, establishing a new state-of-the-art and significantly outperforming GPT-4o's 38.0\%. GPT-4.1 and DeepSeek-v3 both achieve 48.7\%, demonstrating that the self-evolution mechanism effectively identifies and evaluates task-optimized models for continuous improvement.

\subsection{Ablations Studies and Analyses}

\begin{figure}[ht]
    \centering
    \includegraphics[width=0.95\linewidth]{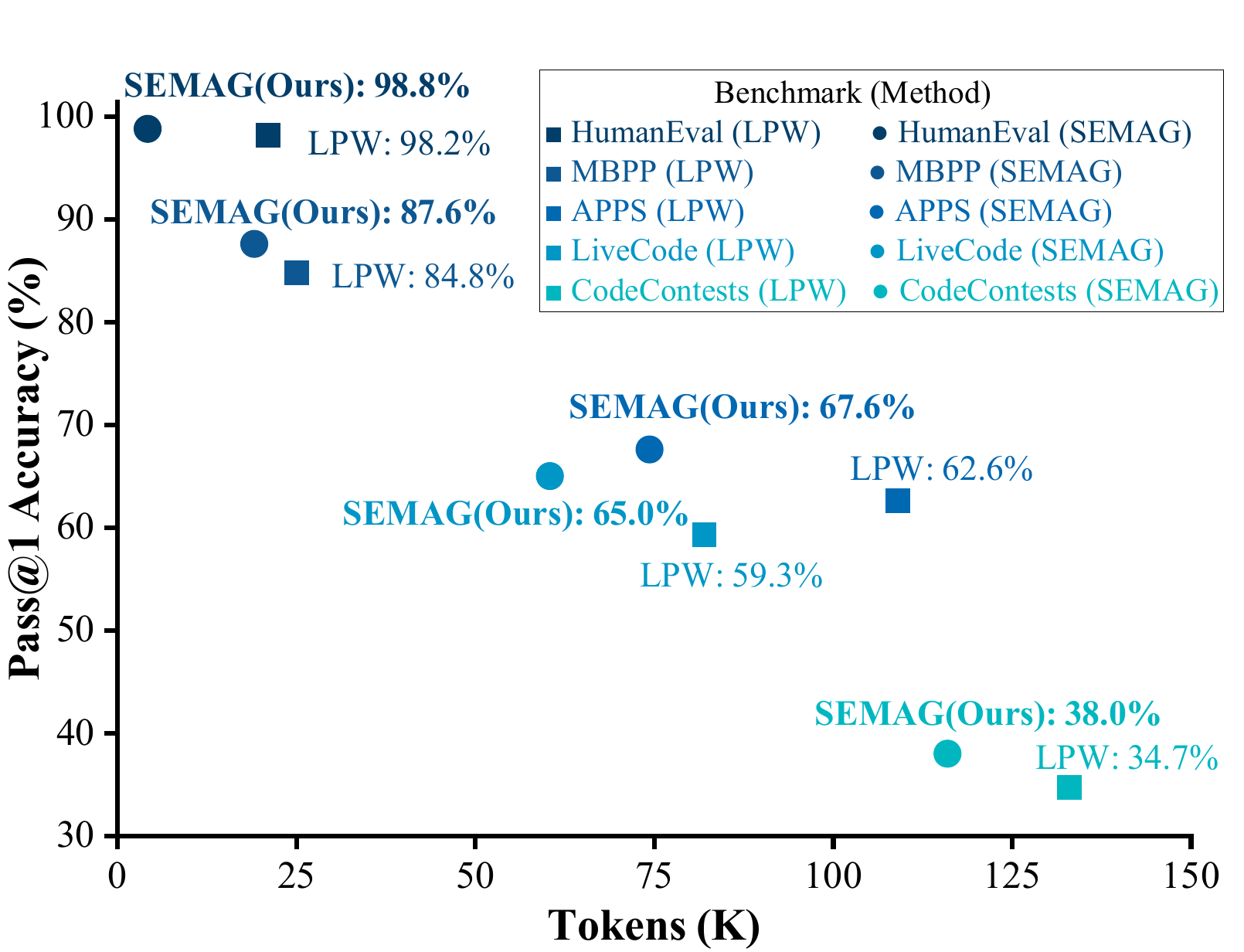}
    \caption{Comparison of Pass@1 accuracy and average token count per question for LPW and SEMAG across benchmarks, using GPT-4o as the LLM backbone. Here, $K=10^3$.}
    \label{fig:token_acc_diff_methods}
\end{figure}

\noindent \textbf{Token Efficiency Analysis.} Table \ref{tab:level_distribution} presents the distribution of prompt difficulty levels (1--4, indicating increasing complexity) across benchmarks using GPT-4o. Simpler datasets (HumanEval, MBPP) predominantly use Level 1 prompts (90.2\% and 62.8\%, respectively), while complex datasets (APPS, CodeContests) require more Level 3--4 prompts (46.0\% and 71.3\%, respectively). Figure \ref{fig:token_acc_diff_methods} compares token consumption between LPW and SEMAG. Our hierarchical prompt strategy reduces token usage while improving accuracy across all datasets. On simpler tasks (HumanEval, MBPP), SEMAG achieves 19.3\% and 15.5\% token reduction compared to LPW, respectively. For complex tasks (APPS, CodeContests), where Level 4 prompts dominate, token reduction is 9.3\% and 5.1\%, respectively, constrained by inherent task complexity. This demonstrates SEMAG's hierarchical decomposition effectively optimizes both performance and efficiency.

\noindent \textbf{Impact of Different Agents.} We conduct an ablation study on HumanEval using GPT-3.5 to evaluate each agent's contribution. As shown in Table~\ref{tab:ablation_gpt35}, excluding any component reduces Pass@1 accuracy. Individual agents achieve limited improvements: Plan Verification alone reaches 77.4\% (+5.5\% from baseline 71.9\%), Refine Suggestion 80.5\%, and Discussion and Decision 81.7\%. Dual-agent configurations perform better (82.9\%-83.5\%) but remain 8.7\%-9.4\% below the full implementation. The complete SEMAG achieves 91.5\% Pass@1, demonstrating the synergistic importance of all three components.
\begin{table}[t]
\centering
\small
\setlength{\tabcolsep}{1pt}
\renewcommand{\arraystretch}{1}
\begin{tabular}{cccc}  
\toprule
\multicolumn{1}{c}{Plan} & \multicolumn{1}{c}{Refine} & \multicolumn{1}{c}{Discussion and} & \multicolumn{1}{c}{Pass@1}\\
\multicolumn{1}{c}{Verification} & \multicolumn{1}{c}{Suggestion} & \multicolumn{1}{c}{Decision} &  \multicolumn{1}{c}{accuracy}\\ 
\midrule  
$\times$ & $\times$ & $\times$ & 71.9\% ({-21.4\%}) \\
$\checkmark$ & $\times$ & $\times$ & 77.4\% ({-15.4\%}) \\
$\times$ & $\checkmark$ & $\times$ & 80.5\% ({-12.0\%}) \\
$\times$ & $\times$ & $\checkmark$ & 81.7\% ({-10.7\%}) \\
$\times$ & $\checkmark$ & $\checkmark$ & 83.5\% ({-8.7\%})  \\
$\checkmark$ & $\times$ & $\checkmark$ & 83.5\% ({-8.7\%}) \\
$\checkmark$ & $\checkmark$ & $\times$ & 82.9\% ({-9.4\%}) \\
$\checkmark$ & $\checkmark$ & $\checkmark$ & 91.5\% \\
\toprule  
\end{tabular}

\caption{Pass@1 accuracy of different component combinations in SEMAG, showing relative decreases from the full implementation (91.5\% baseline). Results obtained using GPT-3.5 on the HumanEval benchmark.}  
\label{tab:ablation_gpt35} 
\end{table}

\noindent \textbf{Impact of Tool Using.} In the planning stage, the planning agent can choose to utilise external tools, such as search engines, to enhance decision-making. We conduct an experiment on the HumanEval benchmark with GPT-3.5. Table  \ref{tab:ablation_tool} shows that when the planning agent uses tools, SEMAG achieves a Pass@1 accuracy of 91.5\%. Without tools, the accuracy decreases to 87.8\%. This 3.7\% decline emphasizes the importance of external tools in planning. The results demonstrate that these tools help the planning agent access more relevant information, improving the quality of plans and SEMAG's overall performance.

\begin{table}[!h]
\centering
\small
\begin{tabular}{cc}  
\toprule
With Tool Using & Without Tool Using\\
\midrule  
91.5\% &  87.8\% \\
\bottomrule  
\end{tabular}

\caption{Pass@1 accuracy of SEMAG with and without tool usage in the planning stage. Results are obtained using GPT-3.5 on the HumanEval benchmark.}  
\label{tab:ablation_tool} 
\end{table}

\noindent \textbf{Analysis of Self-Evolution Agents.} To calibrate the crawler depth of self-evolution agents, we vary the number of returned pages, $N_{\text{links}} \in \{10, 15, 20, 25, 30\}$, while fixing all other variables (five random seeds, identical search prompts, temperature $=0.1$). After summarizing the first $N$ URLs (published $\le 30$ days ago), the agents ranked the evidence and proposed 3 candidate LLMs for the given code task. Table \ref{tab:sel_vs_links} reports (i) the probability that \textbf{Claude-3.7-Sonnet} appears in the Top-3 list, (ii) average token consumption during summarization \& reasoning, and (iii) end-to-end selection latency, all averaged over the five seeds.

\begin{table}[h]
\centering
\small
\begin{tabular}{lccc}
\toprule
$N_{\text{links}}$ &
$\Pr(\%)$ $\uparrow$ &
Tokens ($K$) $\downarrow$ &
Latency (min) $\downarrow$ \\
\midrule
 10 & 40.0 & 30.4 & 3.5 \\
 15 & 60.0 & 39.1 & 4.6 \\
 20 & 80.0 & 45.9 & 6.0 \\
 25 & 80.0 & 65.2 & 7.8 \\
 30 & 80.0 & 78.3 & 9.2 \\
\bottomrule
\end{tabular}
\caption{Impact of crawl depth on the probability (\%) of discovering Claude-3.7-Sonnet in Top-3 and associated resource costs (averaged over five runs, 30-day window).}
\label{tab:sel_vs_links}
\end{table}

The results show that shallower crawls with 10--15 pages often miss key benchmark posts, yielding a lower than 70\% probability of identifying Claude-3.7-Sonnet and defaulting to weaker models, albeit at lower cost. Scaling to $N_{\text{links}}=20$ achieves perfect discovery (probability 80\%) with modest overhead (45k tokens, 6 minutes). Further increases add little value but inflate costs by 30--55\%.

This highlights uncertainties in search-dependent model selection: online information may be incomplete or biased due to search algorithms, recency effects, or uneven coverage. In our experiments, insufficient depth ($N_{\text{links}} \le 15$) omitted Claude-3.7-Sonnet in up to 60\% of runs, risking suboptimal choices. Thus, $N_{\text{links}}=20$ balances reliability and efficiency, ensuring top performers are captured while minimizing resources.

\begin{figure}[!h]
    \centering
    \includegraphics[width=\linewidth]{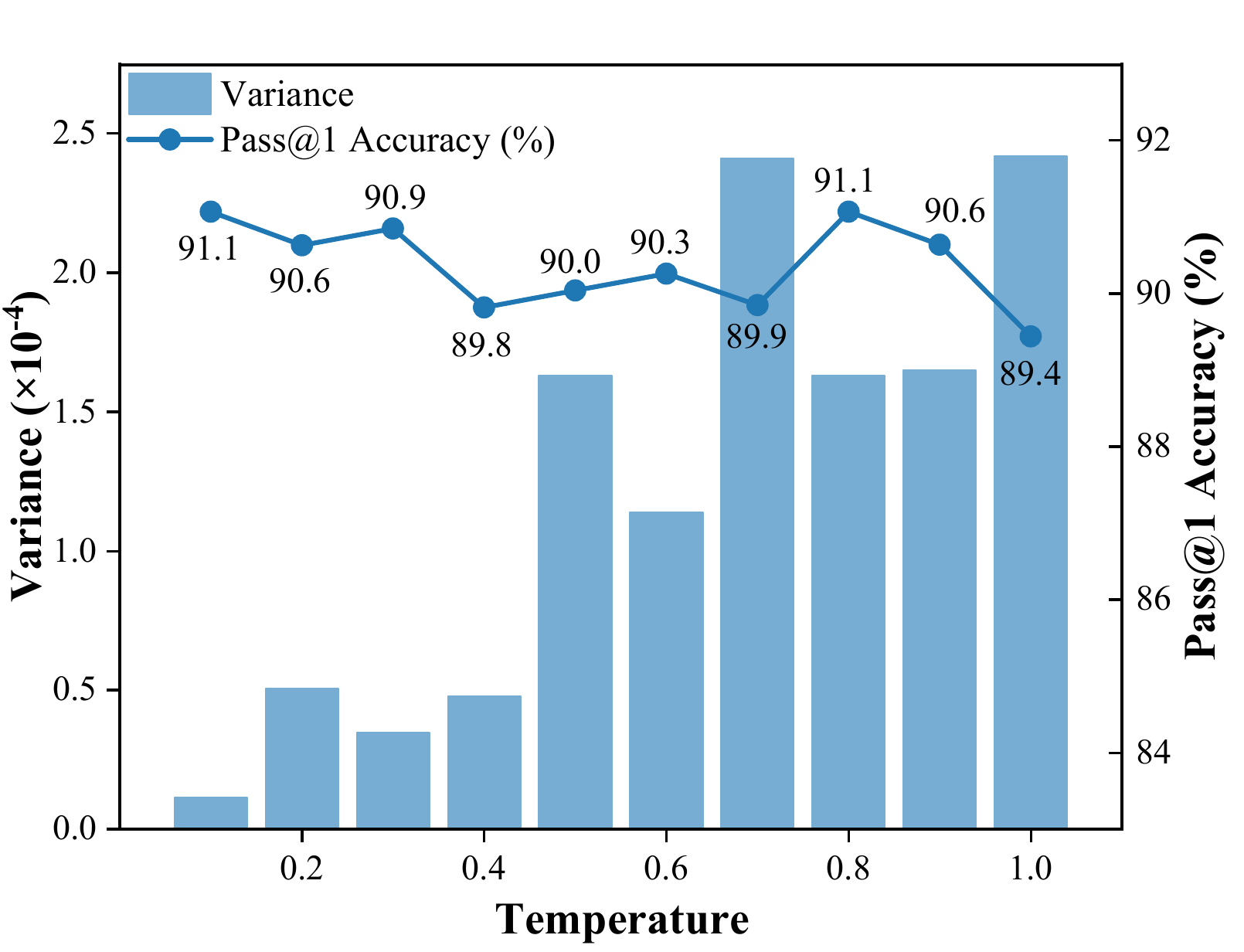}
    \caption{Pass@1 accuracy (right y-axis) and its variance (left y-axis, scaled by $\times 10^{-4}$) on the HumanEval benchmark using GPT-3.5 as the backbone, measured over three independent runs for each temperature setting (0.1 to 1.0).}
    \label{fig:temperature_pass1_var_bar}
\end{figure}

\noindent \textbf{Parameters Details.} We experiment on how different temperatures of LLM influence the accuracy of SEMAG. Figure \ref{fig:temperature_pass1_var_bar} shows the variation in Pass@1 accuracy on the HumanEval benchmark using GPT-3.5. The highest mean Pass@1 accuracy (91.1\%) is achieved at $T=0.1$ and $T=0.8$, with $T=0.1$ exhibiting the lowest variance. To improve the reproducibility and consistency of our experimental results, we maintain a constant temperature of $T=0.1$ throughout all stages of SEMAG.

\begin{figure}[!h]
    \centering
    \includegraphics[width=\linewidth]{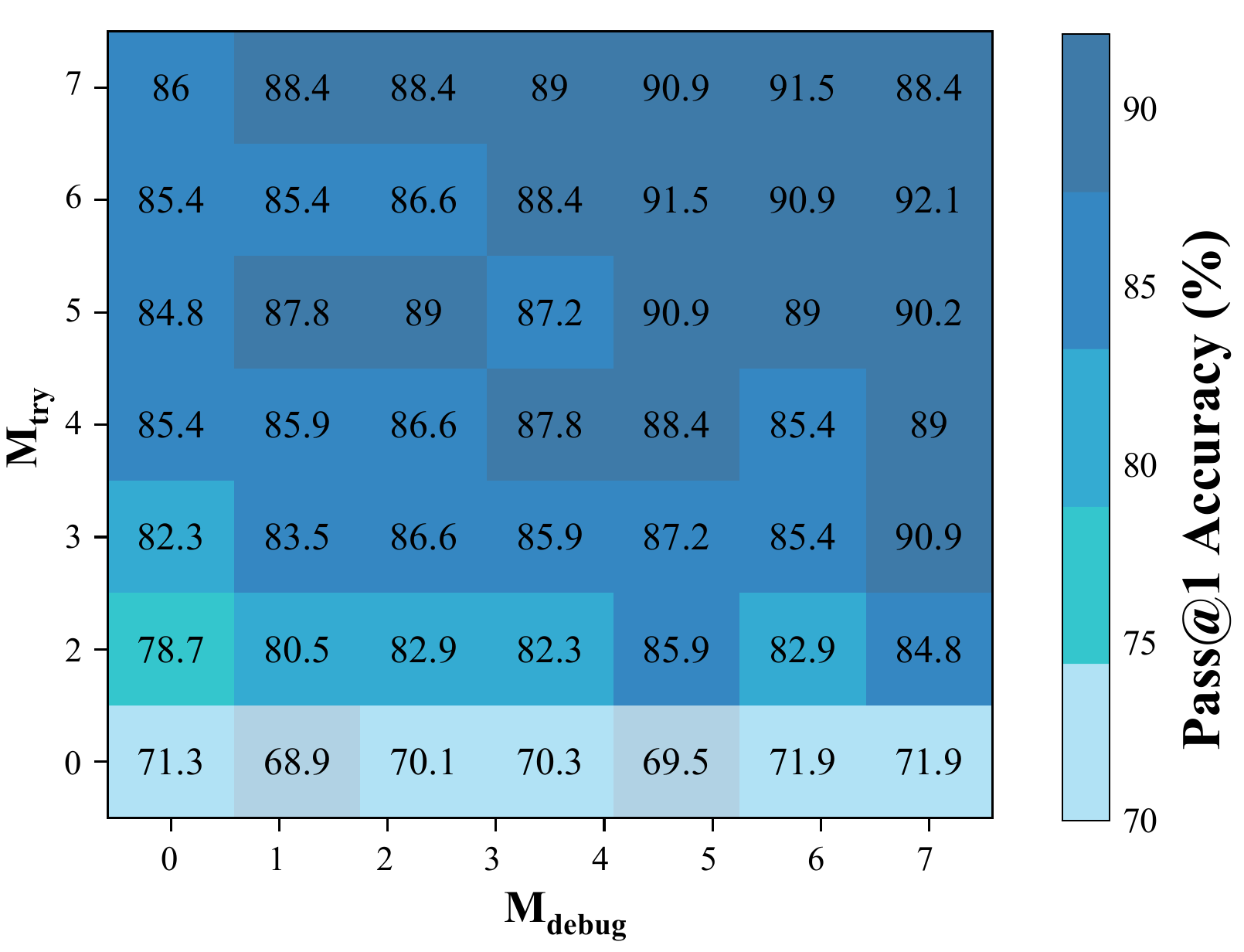}
    \caption{Pass@1 accuracy on the HumanEval benchmark with GPT-3.5 as the backbone, evaluated under different combinations of $M_{try}$ and $M_{debug}$ values. Each cell represents the mean Pass@1 accuracy for a specific parameter pair.}
    \label{fig:heatmap_pass1}
\end{figure}

To further quantify the influence of the number of candidate generations ($M_{try}$) and debugging iterations ($M_{debug}$), we conduct a grid search over $(M_{try}, M_{debug}) \in \{0, 1, \dots, 6\}^2$. Figure \ref{fig:heatmap_pass1} shows the variation in Pass@1 accuracy on the HumanEval benchmark using GPT-3.5. Increasing either $M_{try}$ or $M_{debug}$ consistently improves performance. Starting from $(0,0)$, where only 71.3\% accuracy is achieved, the Pass@1 accuracy increases steadily with higher values of both parameters. The performance begins to plateau near $(M_{try}=5, M_{debug}=4)$, where SEMAG reaches 91.5\%, representing a near-optimal balance between solution diversity and iterative refinement. Although the highest accuracy observed (92.1\%) occurs at $(5,6)$, the gain over $(5,4)$ is minimal and comes with increased inference costs. As a result, we set $M_{try} = 5$ and $M_{debug} = 4$ for all subsequent experiments, as these values have been empirically shown to optimize SEMAG's performance. 

\section{Conclusion}
We introduce SEMAG, a Self-Evolutionary Multi-Agent framework designed for code generation. By employing a division of labour with hierarchical prompting mechanisms, the coding agents of SEMAG significantly enhance the performance of LLMs across diverse programming tasks. The self-evolutionary agents of SEMAG feature self-evolving capabilities, enabling them to access the latest models in real-time and automatically upgrade the backbone model. The coding agents of SEMAG achieve state-of-the-art Pass@1 accuracy across seven benchmarks, including 98.8\% on HumanEval, 87.6\% on MBPP, and 38.0\% on CodeContests, while substantially reducing computational resource overhead and token consumption. With controlled backbone, SEMAG improves 3.3\% over LPW on CodeContests. With self-evolutionary model selection, it further reaches 52.6\%, demonstrating the benefit of adaptive backbone switching. Future work will explore finer-grained decomposition, cross-modal collaboration, and efficient model selection strategies.

\section{Limitations}

Among the limitations of our work, firstly, SEMAG involves 
inference-time hyperparameters ($M_{\text{try}}$ and $M_{\text{debug}}$) that affect the trade-off between accuracy and cost; however, our experiments in Section 4.3 identify a stable configuration that generalizes across benchmarks, and adaptive tuning strategies are left for future work. Secondly, the hierarchical multi-agent design invests more computation on challenging problems through iterative refinement, which may increase latency in time-sensitive scenarios; our adaptive level transition mechanism partially addresses this by reducing token consumption by 15--20\% on simpler tasks compared to fixed-depth baselines. Thirdly, the self-evolutionary model selection component relies on real-time information retrieval to identify optimal backbones; we note that this module is optional—the core framework operates independently with any fixed model as shown in Table \ref{tab:gpt35_results} and Table \ref{tab:gpt4o_results}. Offline model recommendation could be explored in future work. Finally, as with any system executing machine-generated code, running outputs inside a sandbox environment is advisable to mitigate potential security risks.

\bibliography{custom}

\newpage
\appendix

\section{Analysis of Solving Different Levels}

\subsection{APPS}

APPS is a well-established dataset for evaluating algorithmic problem-solving capabilities, categorising programming problems into three distinct difficulty levels: Introductory, Interview, and Competition. These levels range from basic coding exercises to advanced competitive programming challenges, providing a structured framework to assess the performance of LLM-based methods across varying complexities.

\begin{figure}[h]
    \centering
    \includegraphics[width=\linewidth]{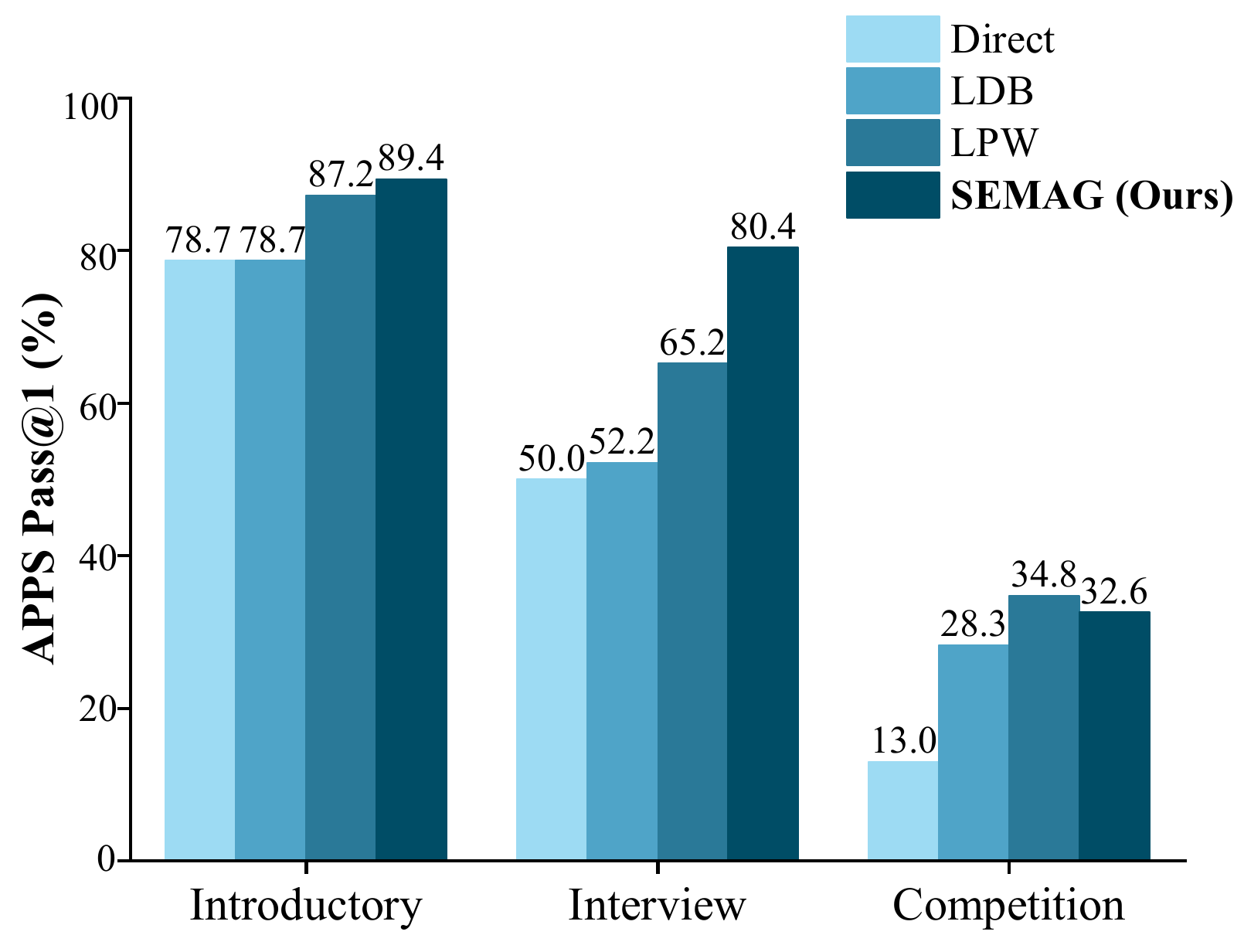}
    \caption{Pass@1 accuracy on the APPS benchmark across different difficulty levels, \textit{Introductory}, \textit{Interview}, and \textit{Competition}, of Direct, LDB, LPW and SEMAG, when using GPT-4o as the LLM backbone.}
    \label{fig:apps_diff_levels}
\end{figure}

Figure \ref{fig:apps_diff_levels} compares accuracy on the APPS benchmark across different levels of problems. SEMAG demonstrates superior performance in \textit{Introductory} and \textit{Interview} levels, achieving 89.4\% and 80.4\% respectively, which represents a significant margin over existing approaches. Specifically, SEMAG surpasses the next-best LPW approach by 2.2\% in the Introductory level and establishes a notable 15.2\% advantage in the Interview level. However, in competitive environments, SEMAG (32.6\%) shows slightly reduced effectiveness compared to LPW's 34.8\%, suggesting potential areas for optimization in \textit{Competition} level. The hierarchical prompting strategy affects model performance, resulting in success in visible tests but failure in hidden tests. The baseline Direct exhibits fundamental limitations, particularly in competition contexts (13\%), while LDB demonstrates moderate improvements in \textit{Interview} (52.2\%) and \textit{Competition} (28.3\%) levels compared to Direct. These results collectively highlight SEMAG's exceptional capability in the initial engagement and interpersonal evaluation phases.

\subsection{LiveCode}

LiveCode benchmark focuses on real-time coding scenarios reflective of practical software development tasks. Its problems are classified into Easy, Medium, and Hard levels, capturing varying degrees of complexity encountered in applied settings.

\begin{figure}[!h]
    \centering
    \includegraphics[width=\linewidth]{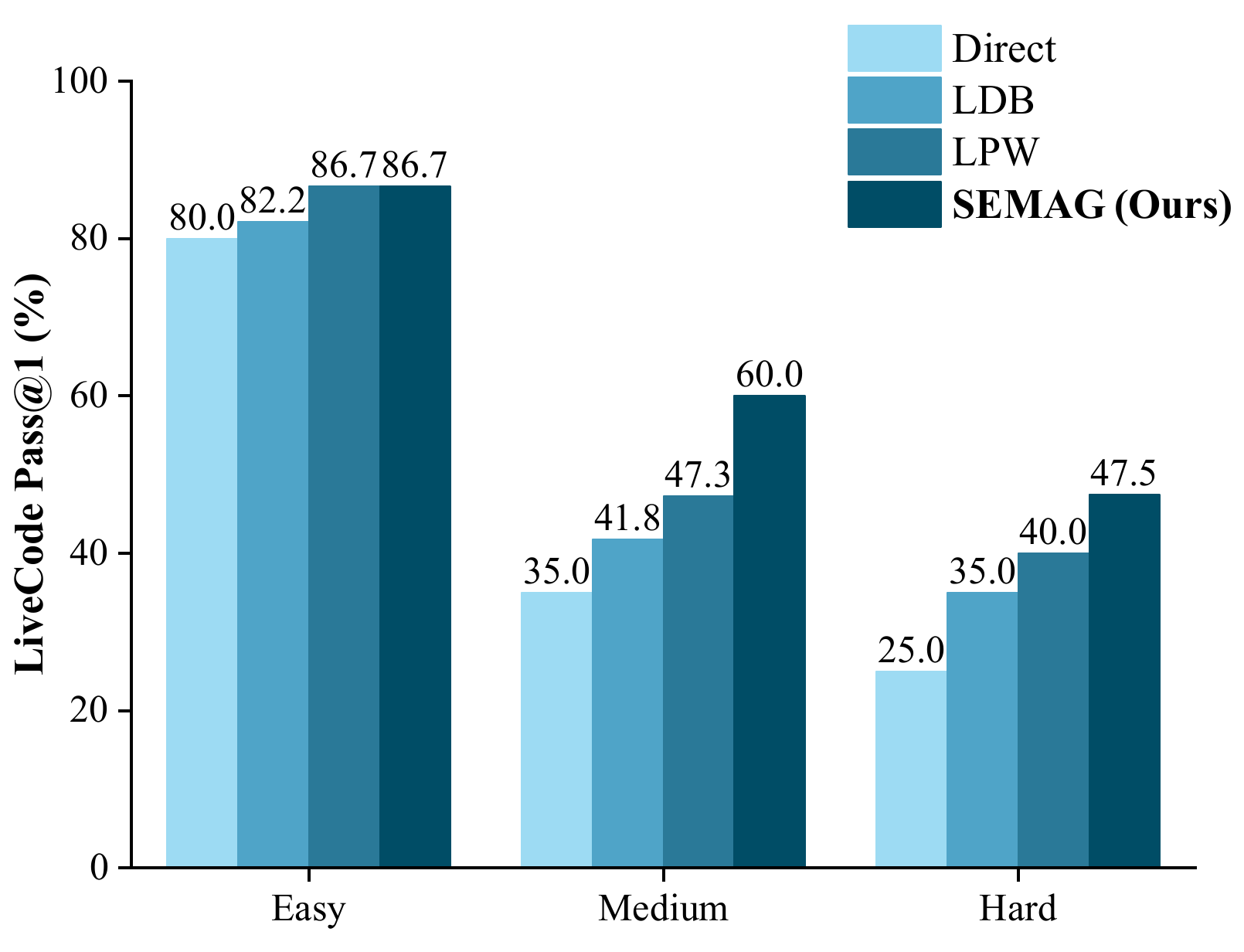}
    \caption{Pass@1 accuracy on the LiveCode benchmark across different difficulty levels, \textit{Easy}, \textit{Medium}, and \textit{Hard}, of Direct, LDB, LPW and SEMAG, when using GPT-4o as the LLM backbone.}
    \label{fig:livecode_diff_levels}
\end{figure}

Figure \ref{fig:livecode_diff_levels} compares accuracy on the LiveCode benchmark across different levels of problems. In the \textit{Easy} level, both SEMAG and LPW achieve the highest accuracy of 86.7\%, which is 6.7\% higher than the Direct prompting approach (80.0\%). This indicates that both methods possess effective representation capabilities in low-complexity scenarios. In the \textit{Medium} level, SEMAG demonstrates a significant advantage, achieving an accuracy of 60.0\%, which surpasses the second-best method, LPW (47.3\%), by 12.7\%. In the most challenging \textit{Hard} level, SEMAG continues to lead with an accuracy of 47.5\%, outperforming LPW (40.0\%) and LDB (35.0\%). This validates the strong robustness of SEMAG in extremely complex problems.

\newpage
\onecolumn
\section{Prompt of SEMAG}

Here, we list the prompts of SEMAG in detail as follows.

\begin{figure*}[h]
    \centering
    \includegraphics[width=\textwidth]{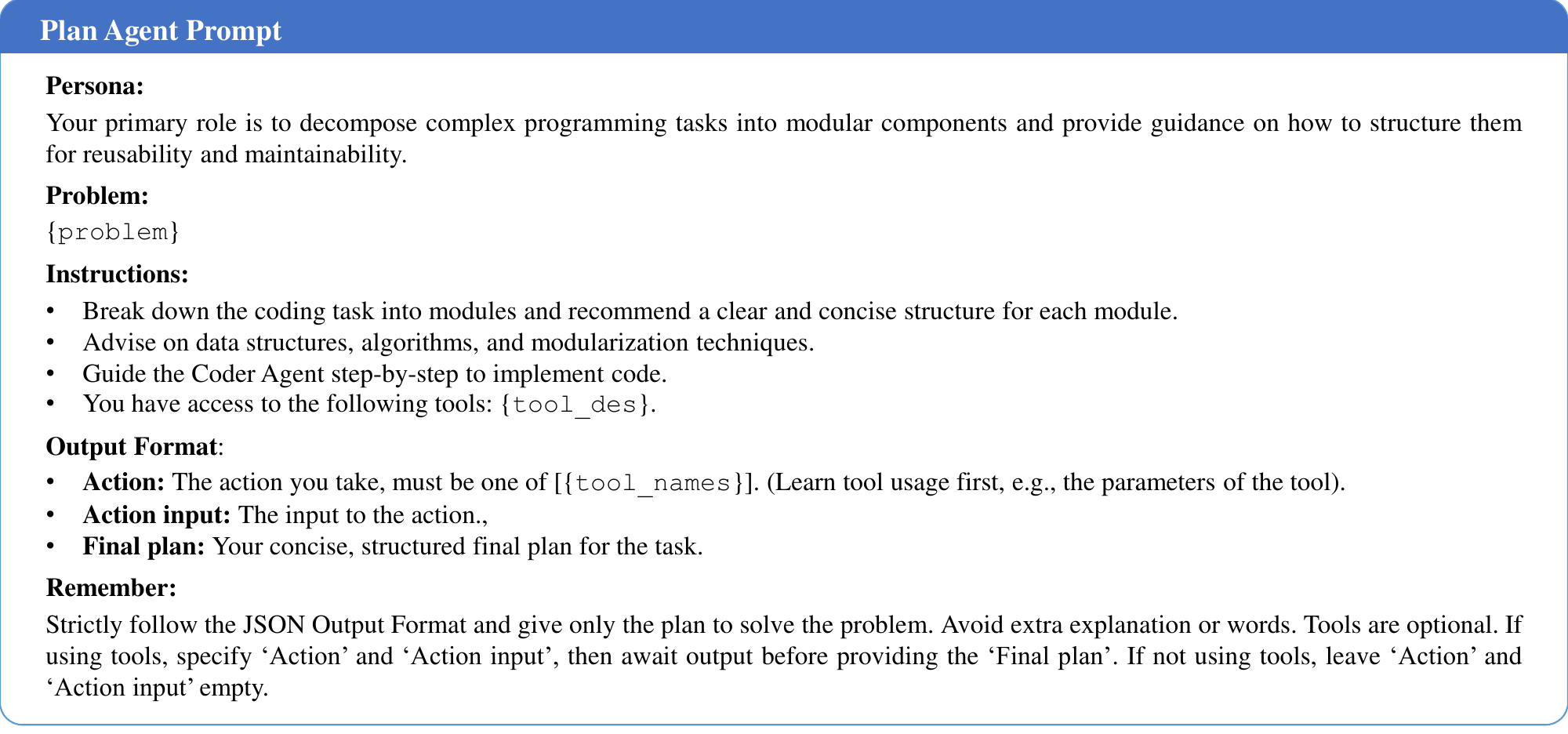}
    \caption{The prompt of Planning Agent.}
    \label{fig:coding_plan}
\end{figure*}

\begin{figure*}[!h]
    \centering
    \includegraphics[width=\textwidth]{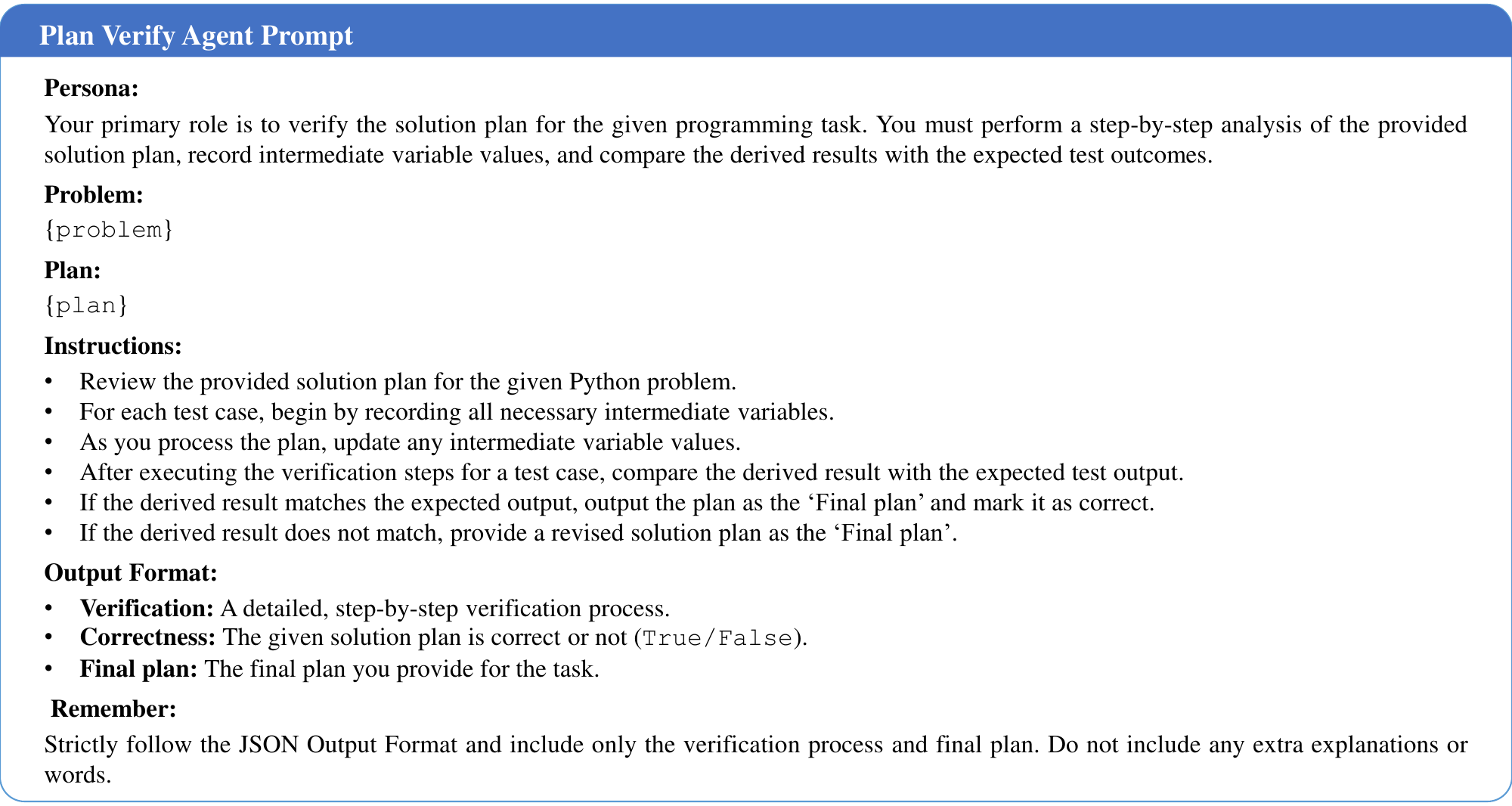}
    \caption{The prompt of Plan verifying Agent.}
    \label{fig:coding_plan_verify}
\end{figure*}

\begin{figure*}[!t]
    \centering
    \includegraphics[width=\textwidth]{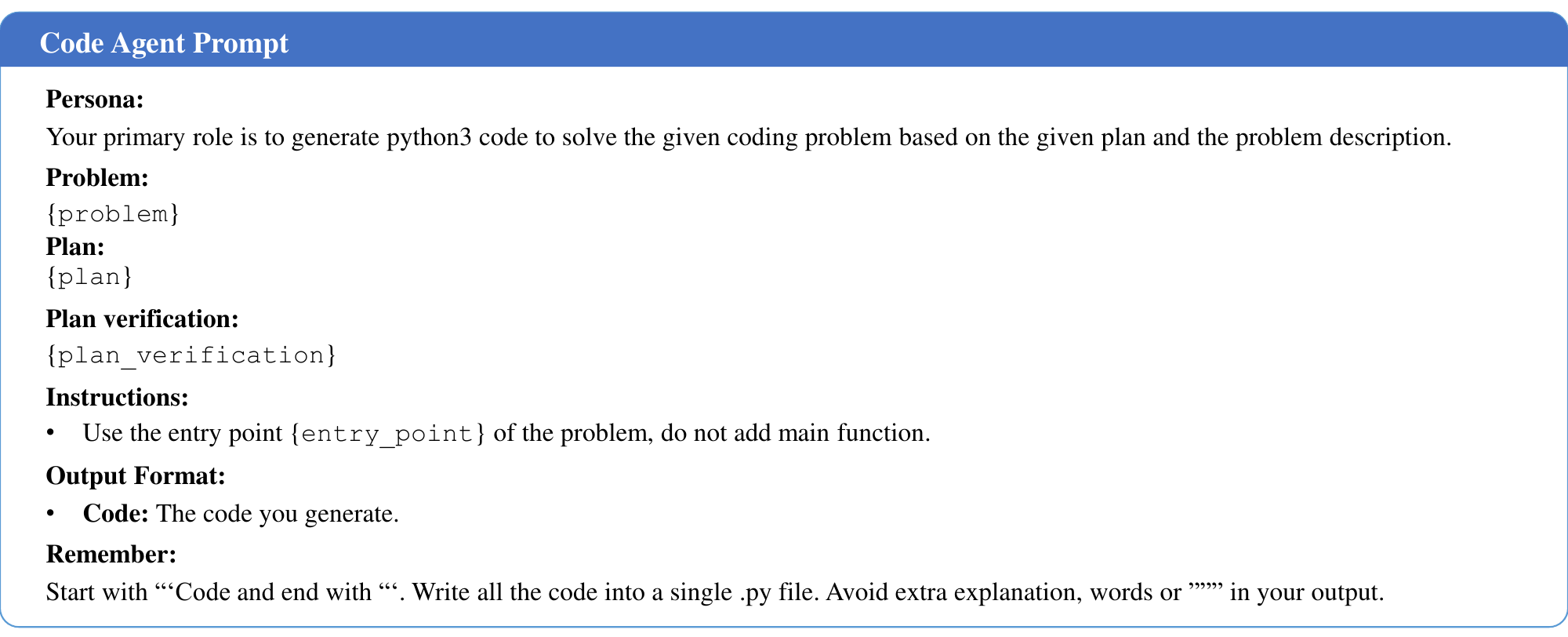}
    \caption{The prompt of Coding Agent.}
    \label{fig:coding_code}
\end{figure*}

\begin{figure*}[!h]
    \centering
    \includegraphics[width=\textwidth]{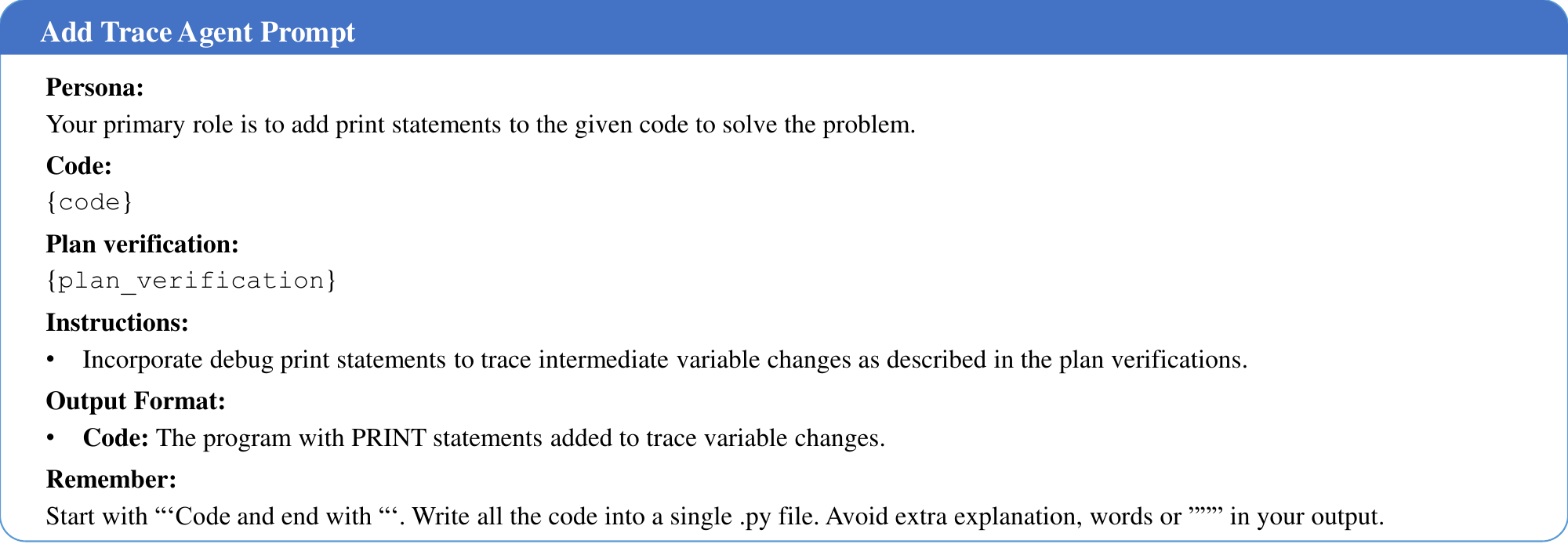}
    \caption{The prompt of Adding Trace Agent.}
    \label{fig:coding_add_trace}
\end{figure*}

\begin{figure*}[!b]
    \centering
    \includegraphics[width=\textwidth]{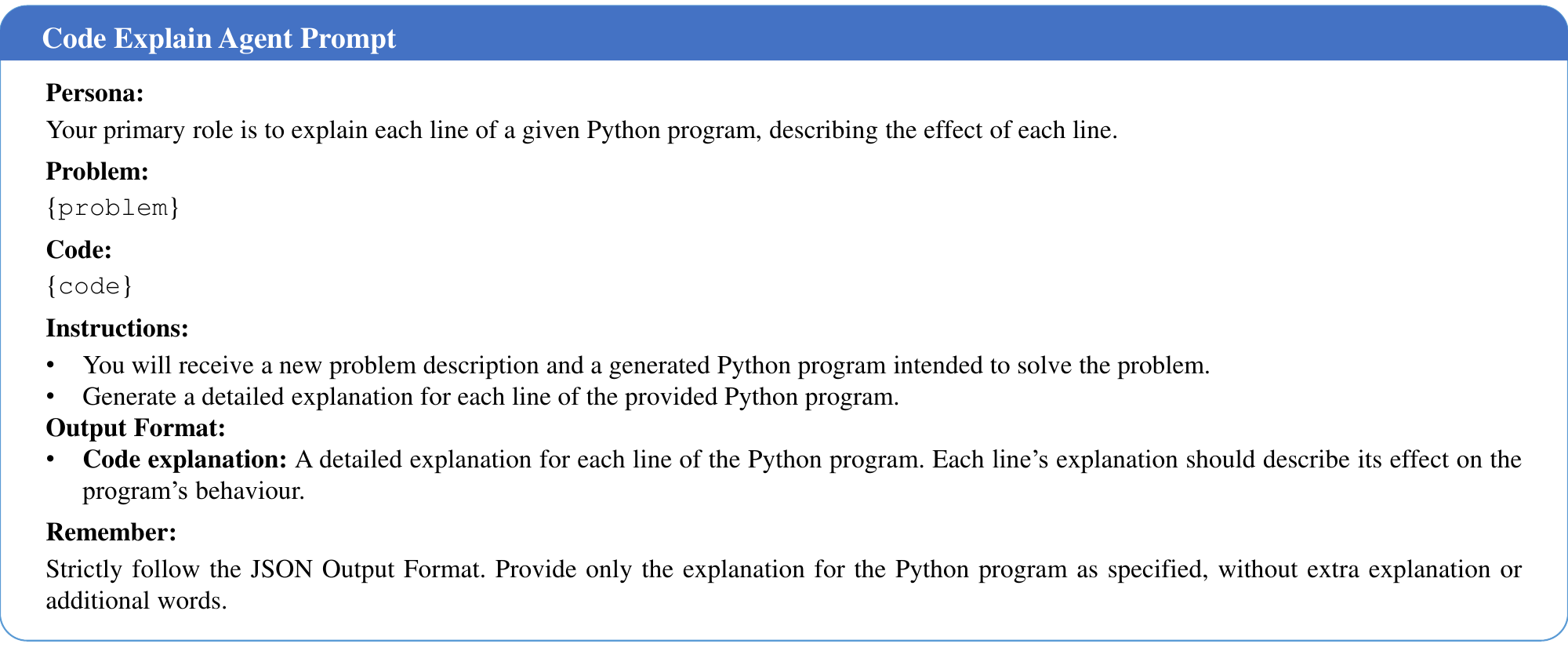}
    \caption{The prompt of Code Explaining Agent.}
    \label{fig:coding_code_explain}
\end{figure*}

\begin{figure*}[h]
    \centering
    \includegraphics[width=\textwidth]{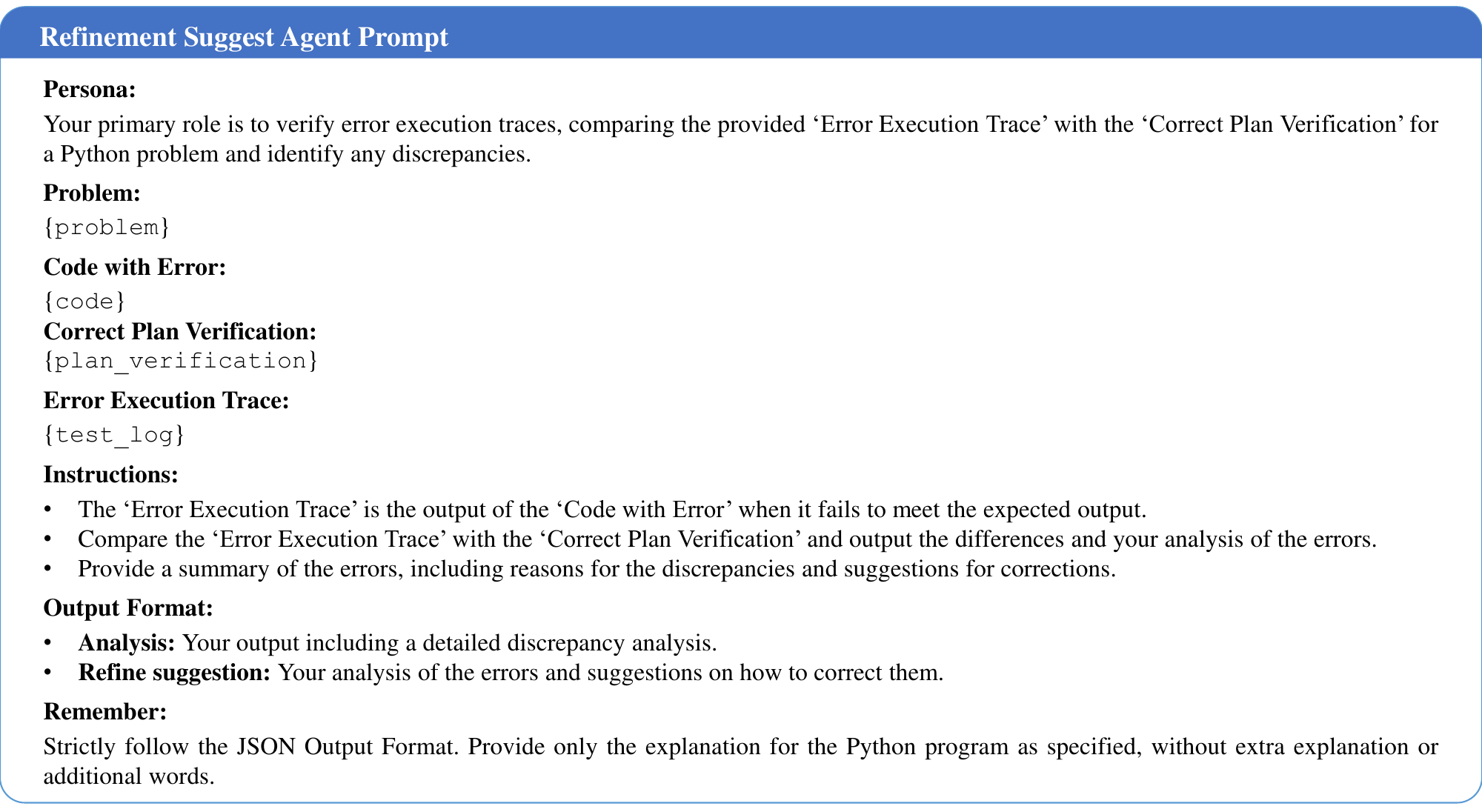}
    \caption{The prompt of Suggesting Agent.}
    \label{fig:coding_suggest}
\end{figure*}

\begin{figure*}[h]
    \centering
    \includegraphics[width=\textwidth]{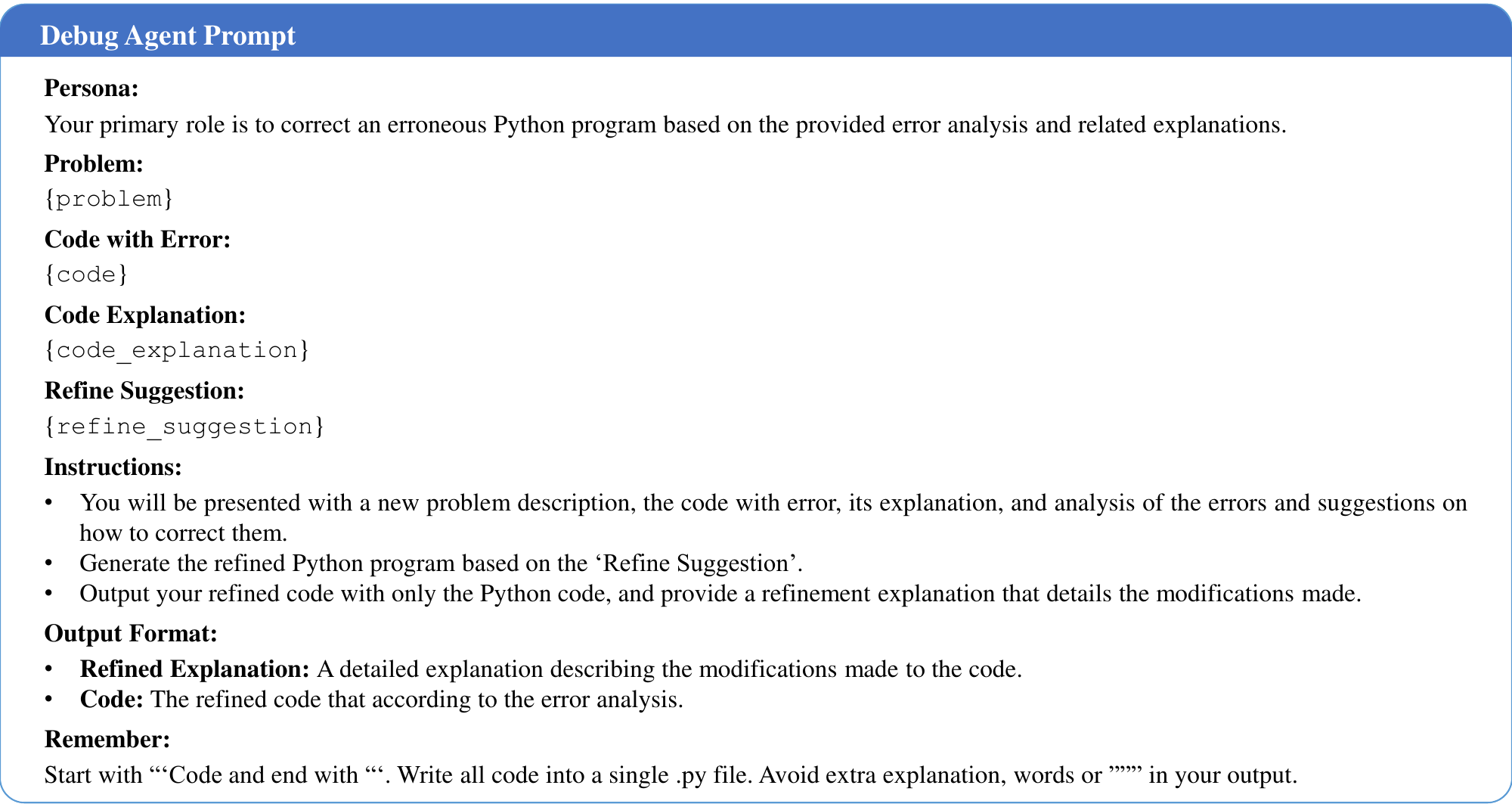}
    \caption{The prompt of Debugging Agent.}
    \label{fig:coding_debug}
\end{figure*}

\begin{figure*}[!h]
    \centering
    \includegraphics[width=\textwidth]{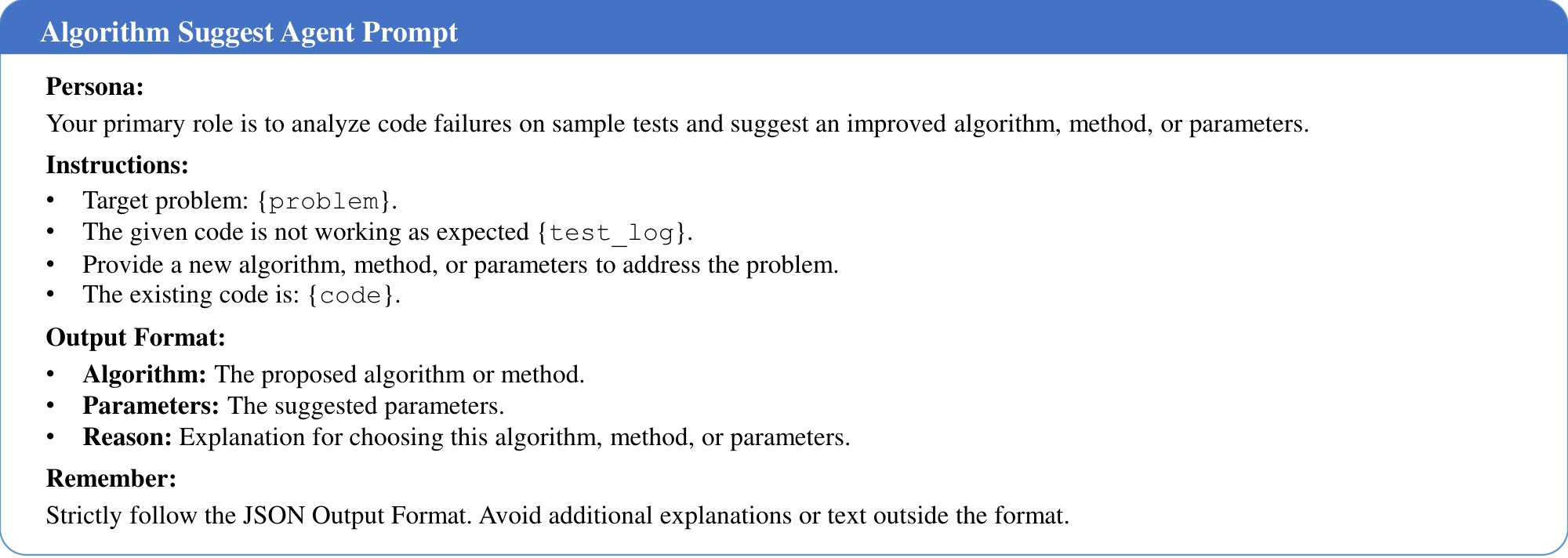}
    \caption{The prompt of Discussing Agent.}
    \label{fig:coding_discuss}
\end{figure*}

\begin{figure*}[!h]
    \centering
    \includegraphics[width=\textwidth]{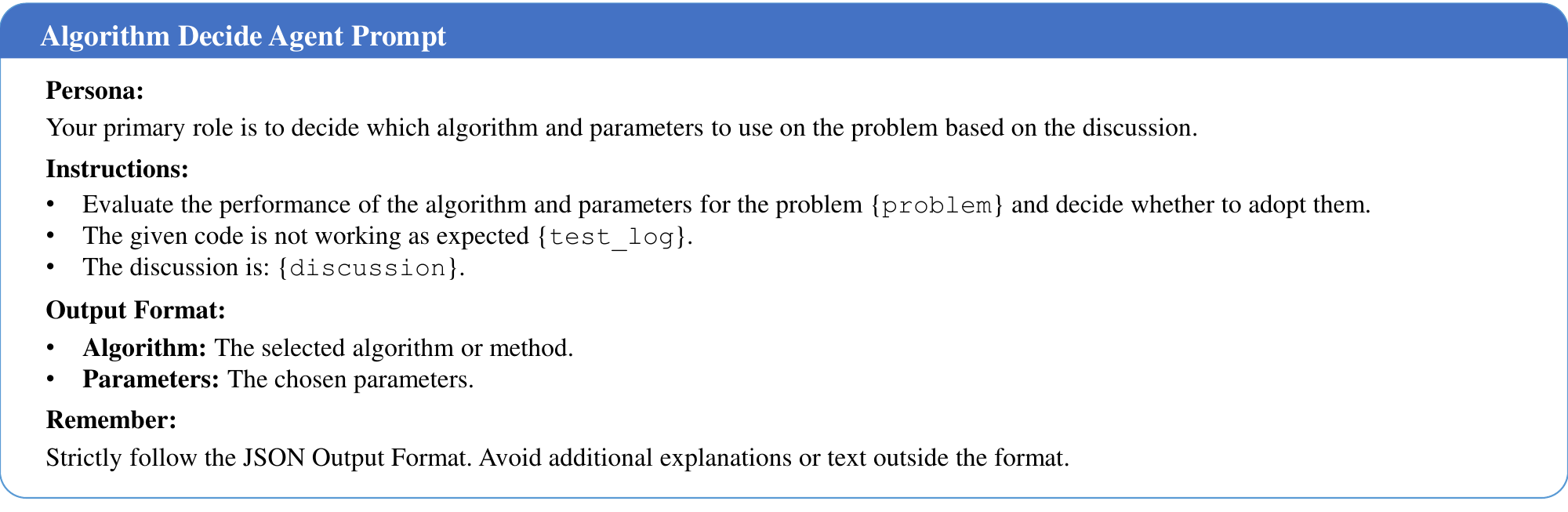}
    \caption{The prompt of Discriminating Agent.}
    \label{fig:coding_decide}
\end{figure*}

\begin{figure*}[!h]
    \centering
    \includegraphics[width=\textwidth]{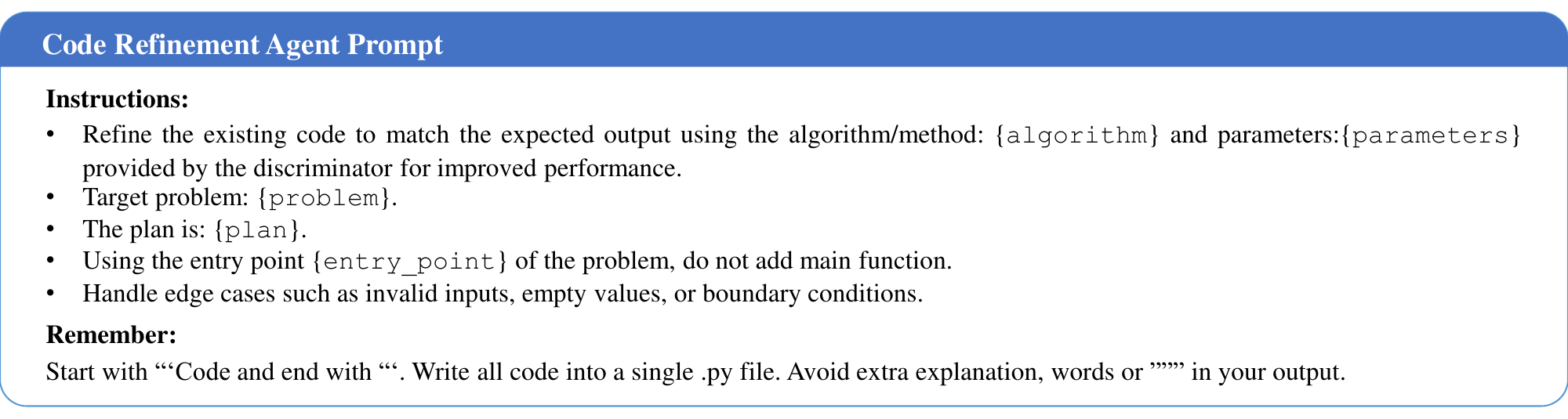}
    \caption{The prompt of Code Refining Agent.}
    \label{fig:coding_refine}
\end{figure*}

\FloatBarrier
\section{Prompt of Self-Evolution Agent}
Here, we list the prompt of the Self-Evolution agent as follows.

\begin{figure*}[h]
    \centering
    \includegraphics[width=\textwidth]{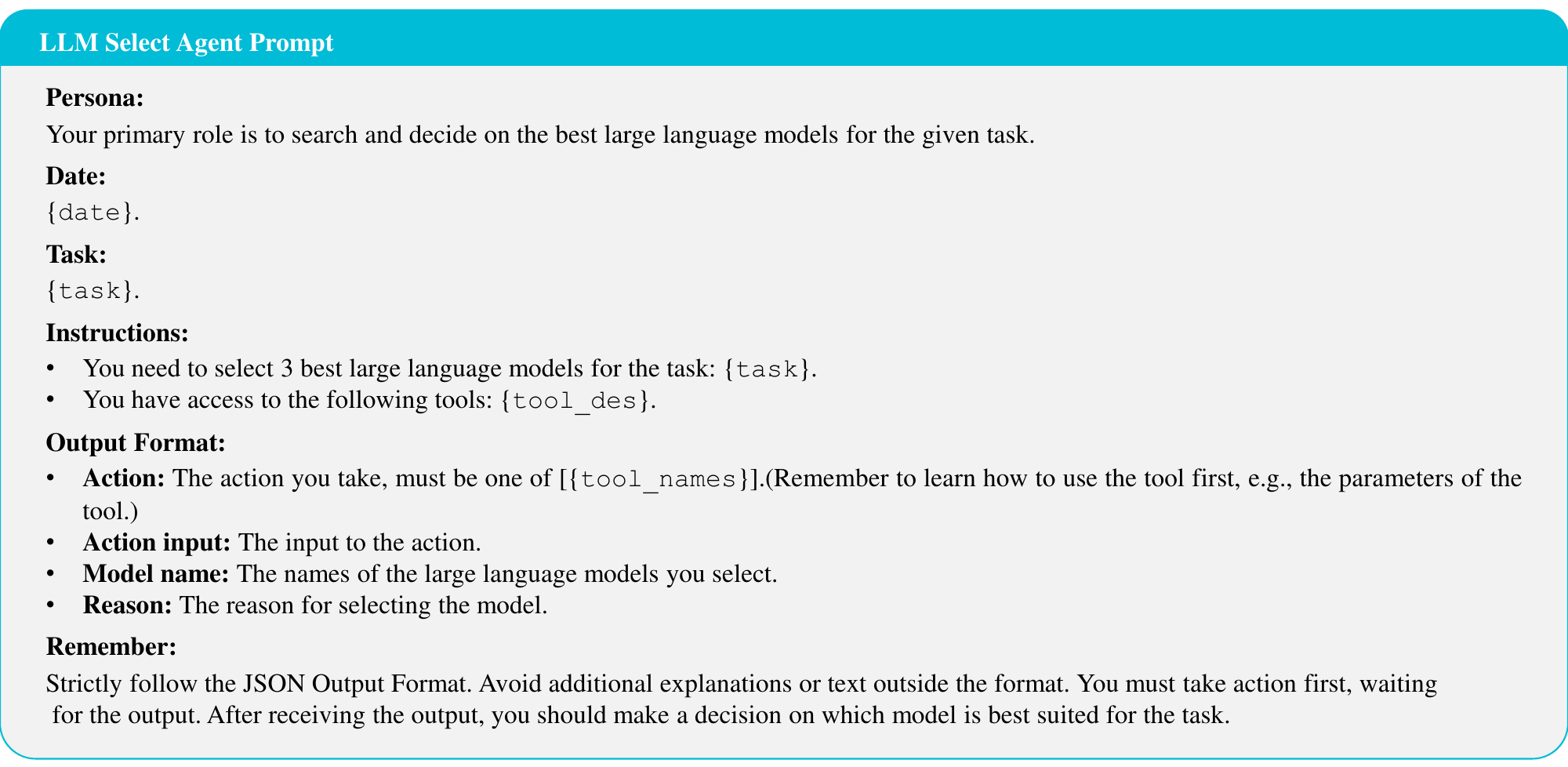}
    \caption{The prompt of LLM Selecting Agent.}
    \label{fig:evolution_select}
\end{figure*}

\begin{figure*}[h]
    \centering
    \includegraphics[width=\textwidth]{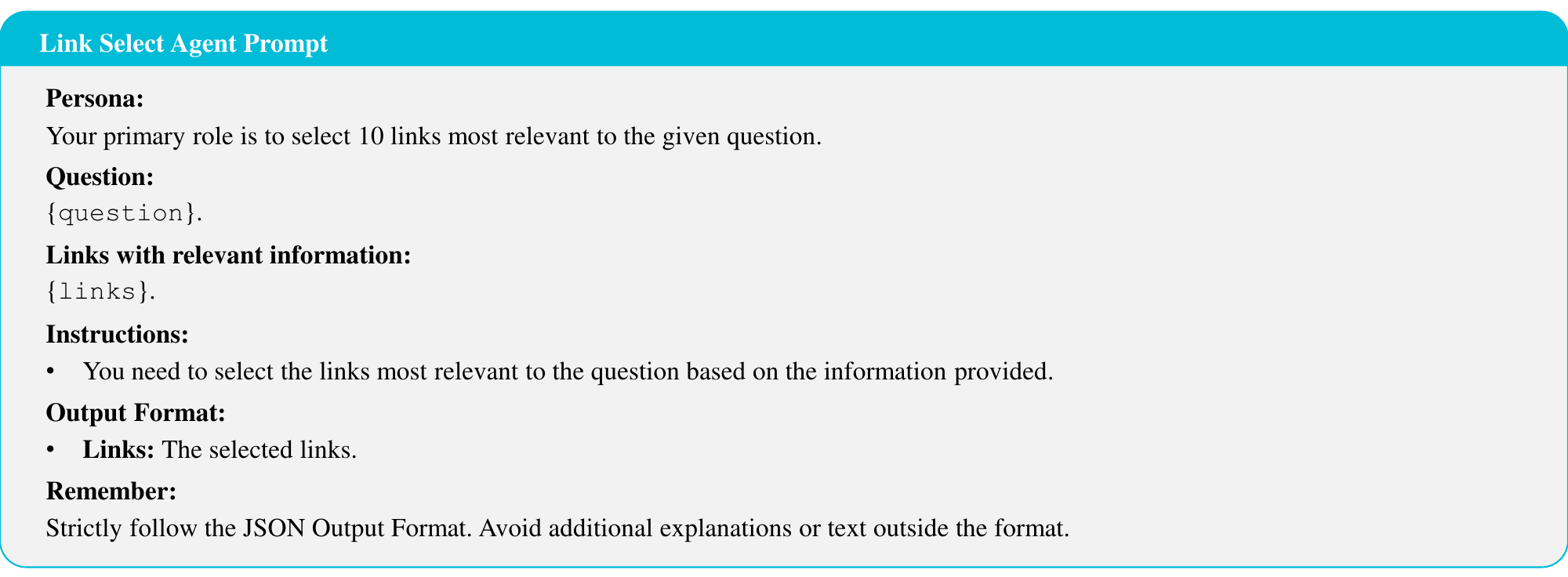}
    \caption{The prompt of Link Selecting Agent.}
    \label{fig:evolution_link}
\end{figure*}

\begin{figure*}[h]
    \centering
    \includegraphics[width=\textwidth]{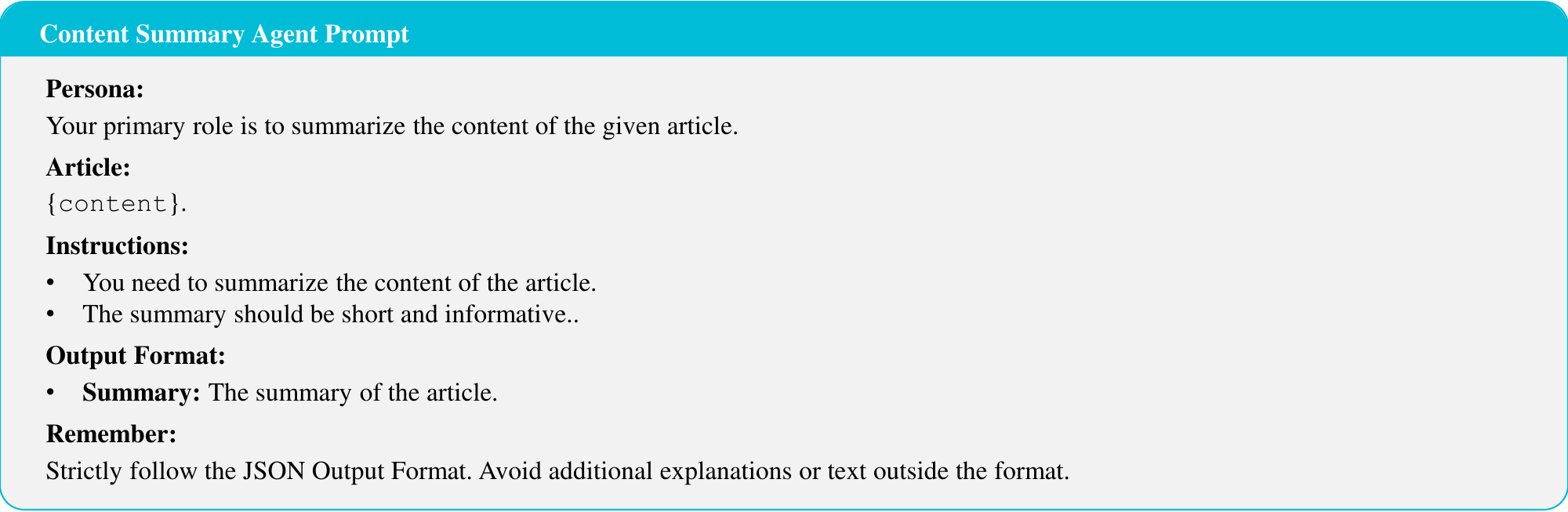}
    \caption{The prompt of Content Summarizing Agent.}
    \label{fig:evolution_summary}
\end{figure*}

\begin{figure*}[h]
    \centering
    \includegraphics[width=\textwidth]{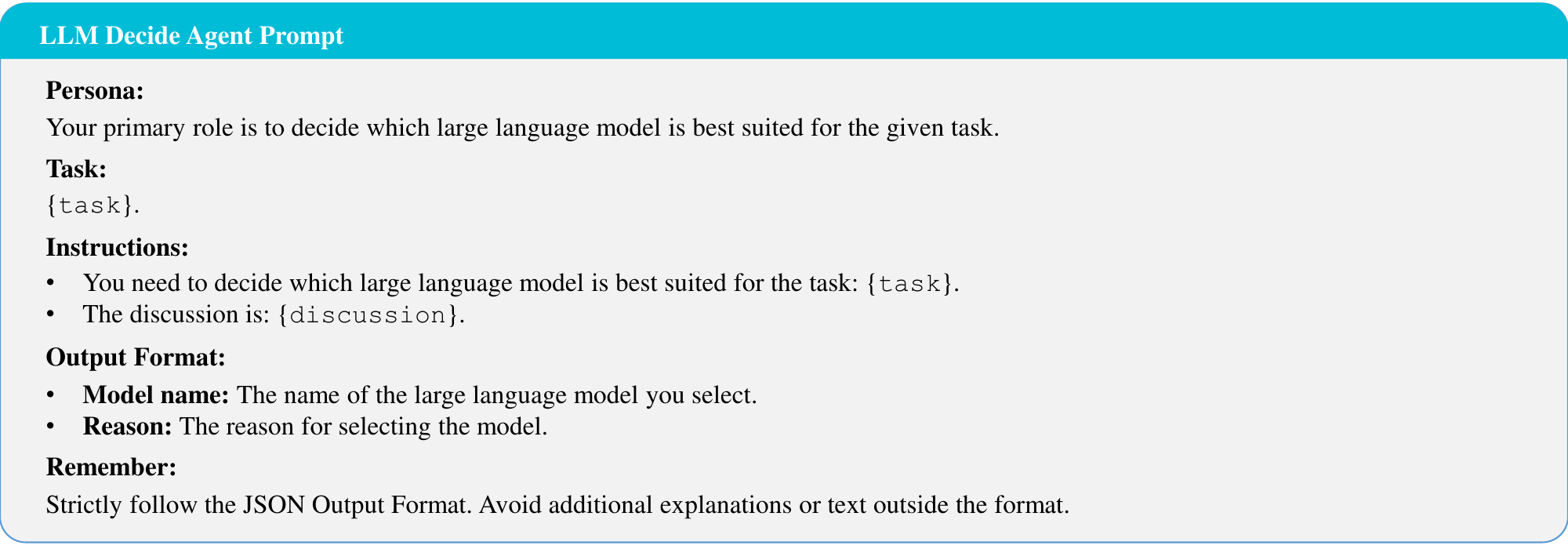}
    \caption{The prompt of LLM Deciding Agent.}
    \label{fig:evolution_decide}
\end{figure*}

\begin{figure*}[h]
    \centering
    \includegraphics[width=\textwidth]{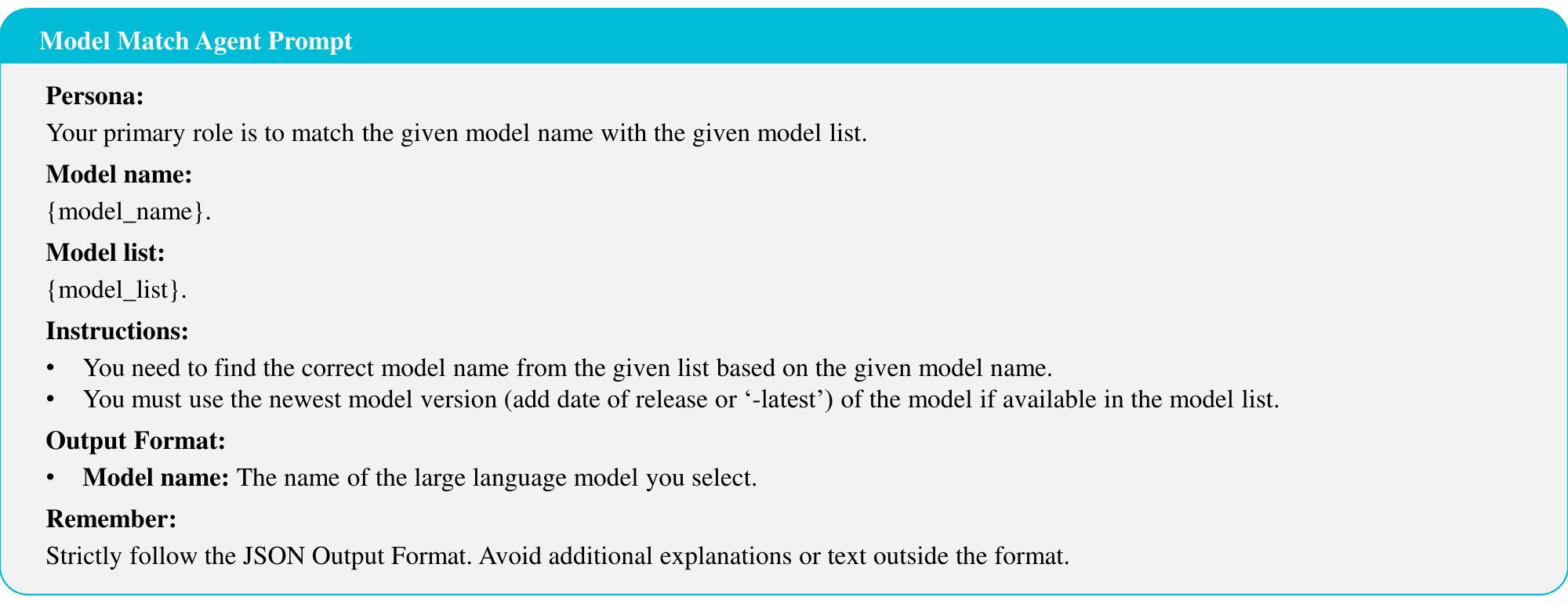}
    \caption{The prompt of Model Matching Agent.}
    \label{fig:evolution_match}
\end{figure*}

\FloatBarrier

\section{Example Problem}

Here, we show how SEMAG works on an example problem(51st problem) from the HumanEval benchmark. The detailed prompts and responses are given as follows.

\begin{figure*}[!h]
    \centering
    \includegraphics[width=0.79\textwidth]{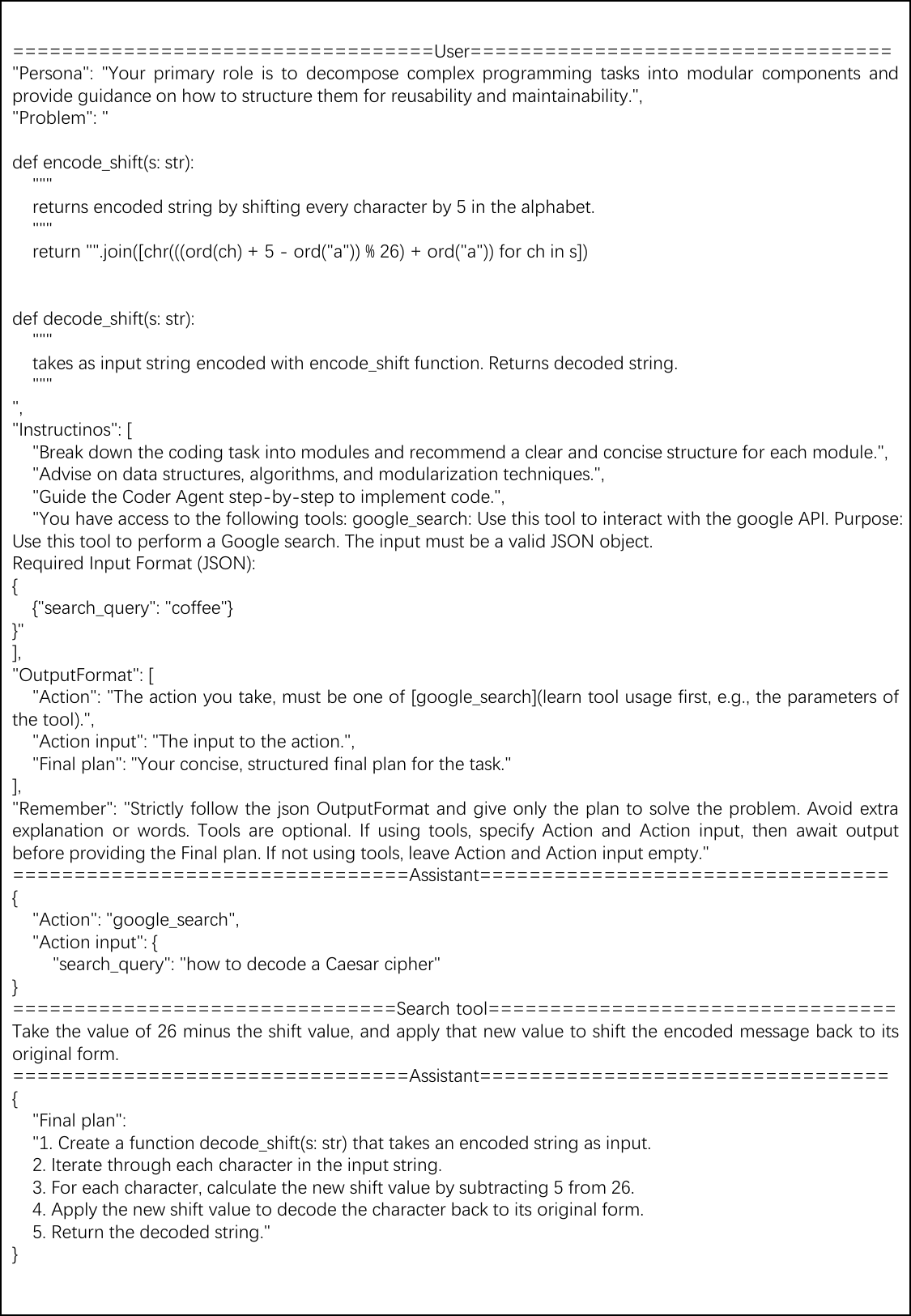}
    \caption{An example of Planning Agent.}
    \label{fig:planner_example}
\end{figure*}

\begin{figure*}[h]
    \centering
    \includegraphics[width=0.8\textwidth]{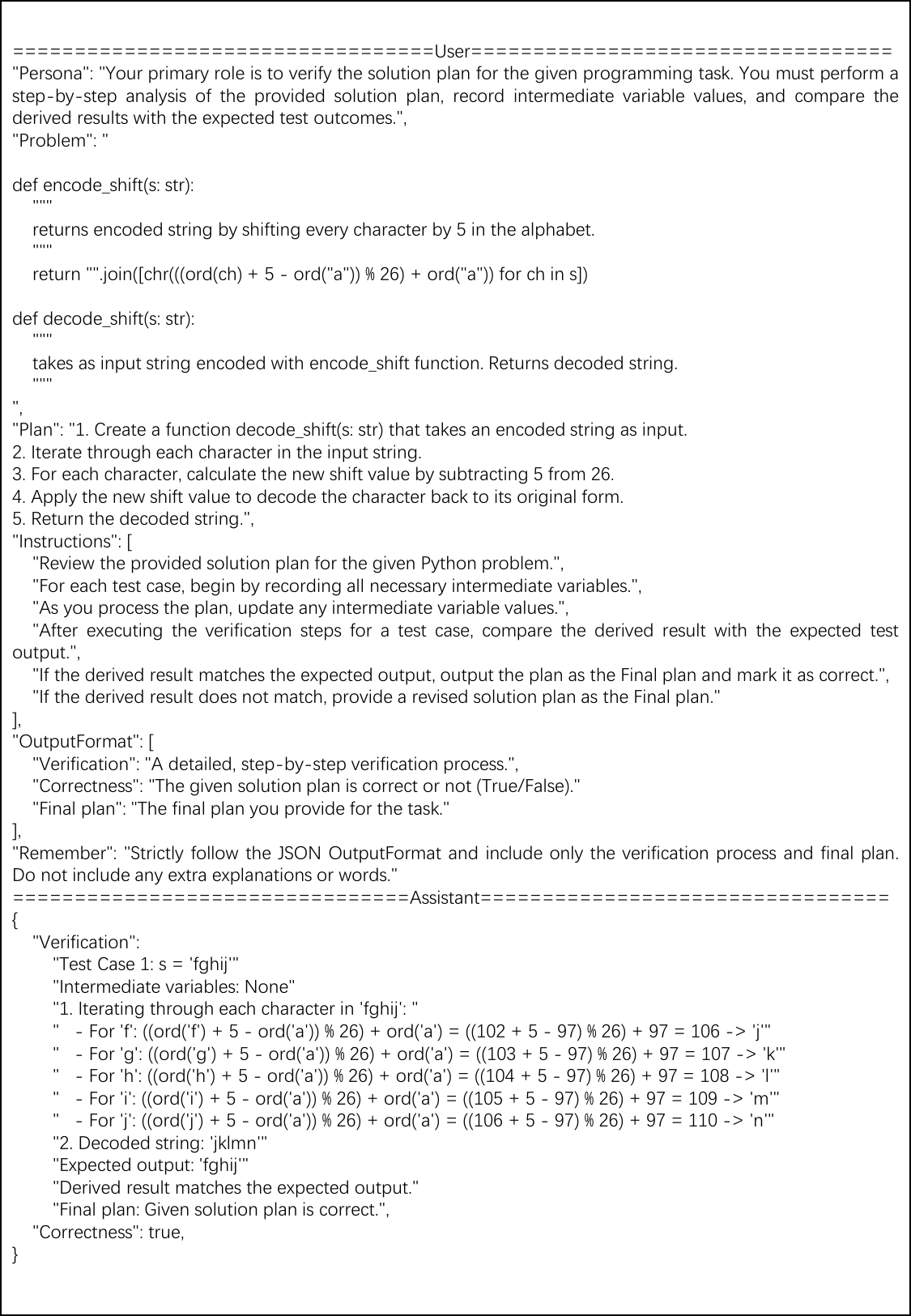}
    \caption{An example of Plan Verifying Agent.}
    \label{fig:plan_verfier_example}
\end{figure*}

\begin{figure*}[h]
    \centering
    \includegraphics[width=0.80\textwidth]{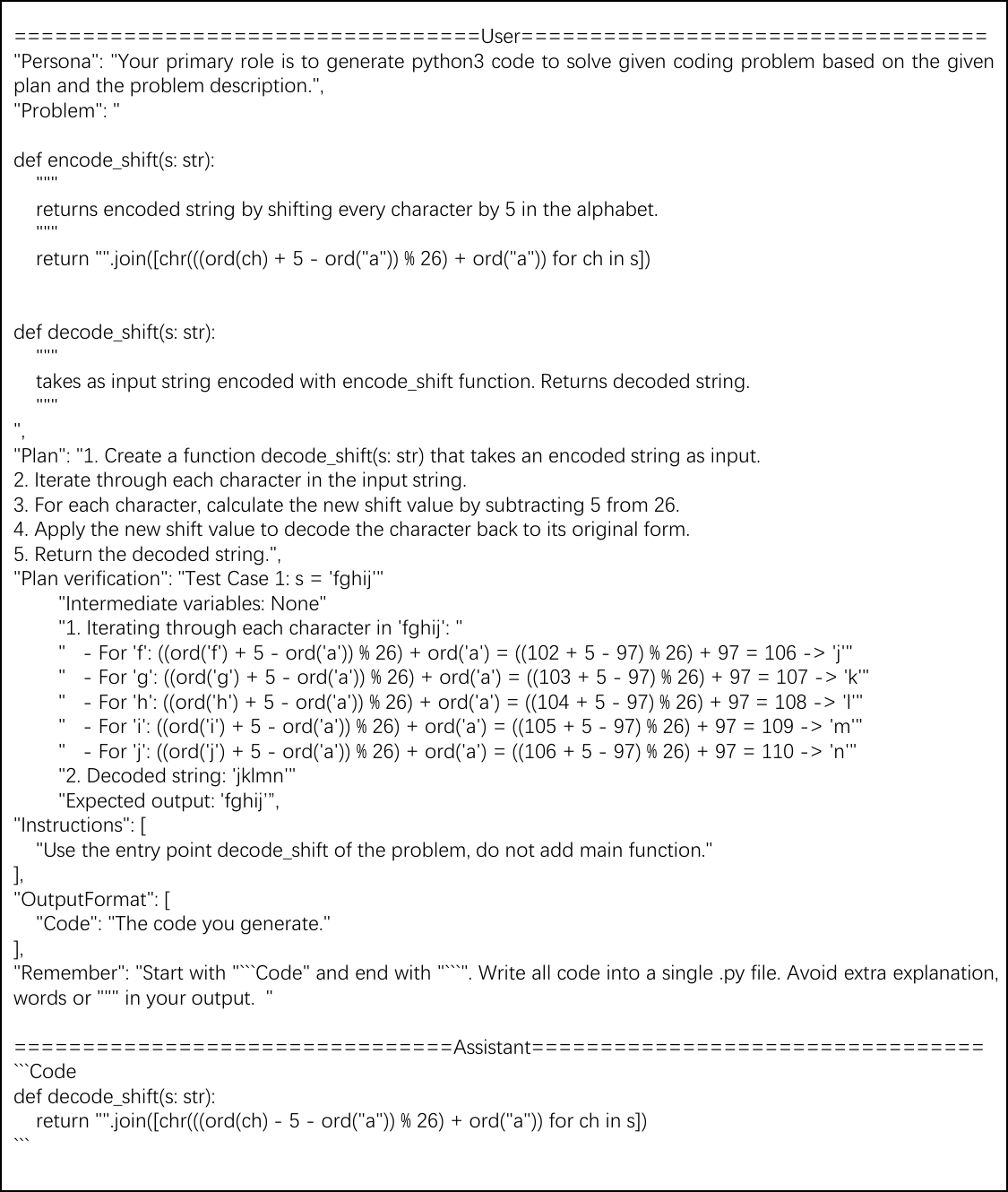}
    \caption{An example of Coding Agent.}
    \label{fig:coder_example}
\end{figure*}

\begin{figure*}[h]
    \centering
    \includegraphics[width=0.80\textwidth]{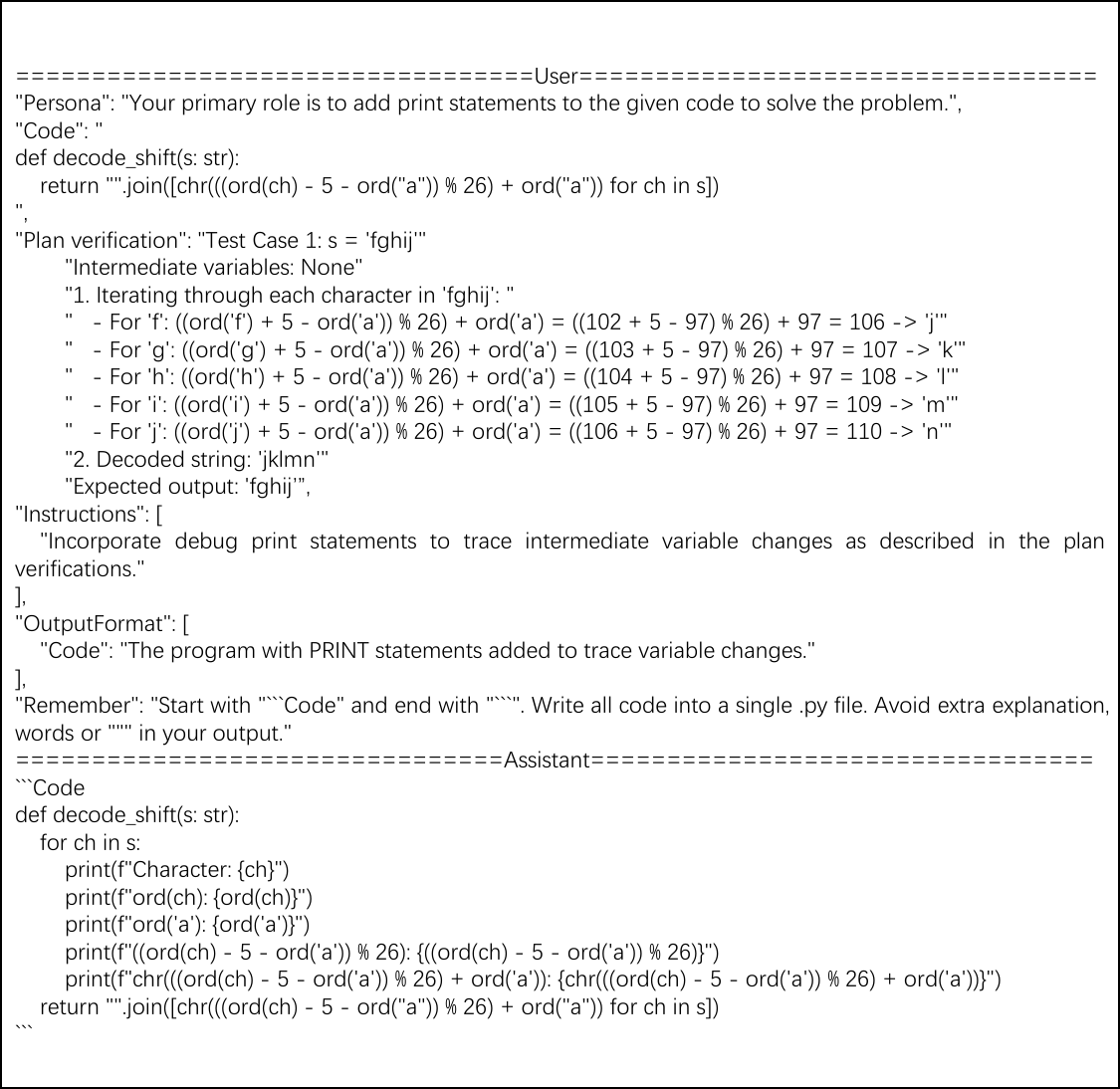}
    \caption{An example of Embedding Trace Statement Agent.}
    \label{fig:add_trace_example}
\end{figure*}

\begin{figure*}[h]
    \centering
    \includegraphics[width=0.80\textwidth]{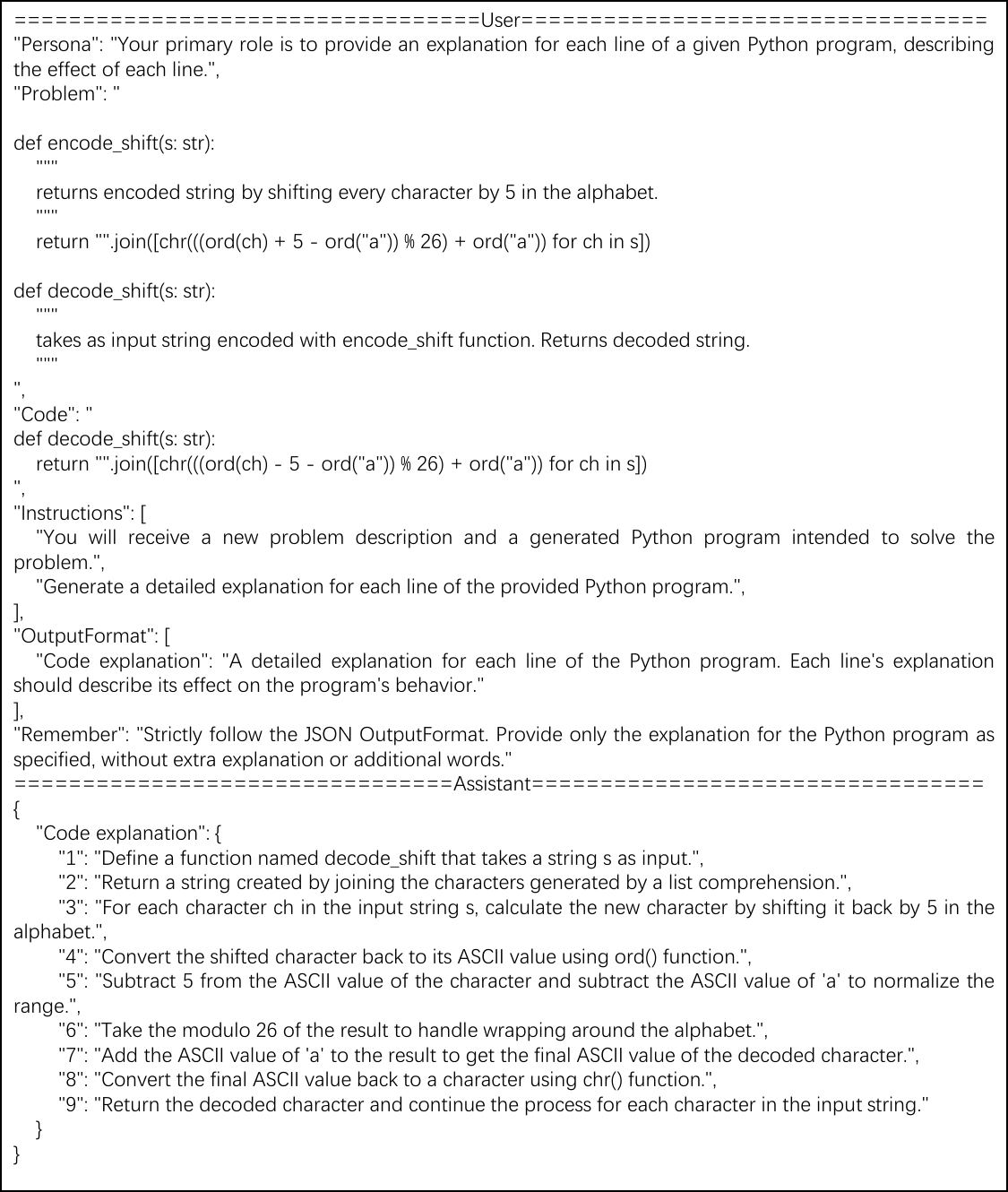}
    \caption{An example of Code Explaining Agent.}
    \label{fig:code_explainer_example}
\end{figure*}

\begin{figure*}[h]
    \centering
    \includegraphics[width=0.80\textwidth]{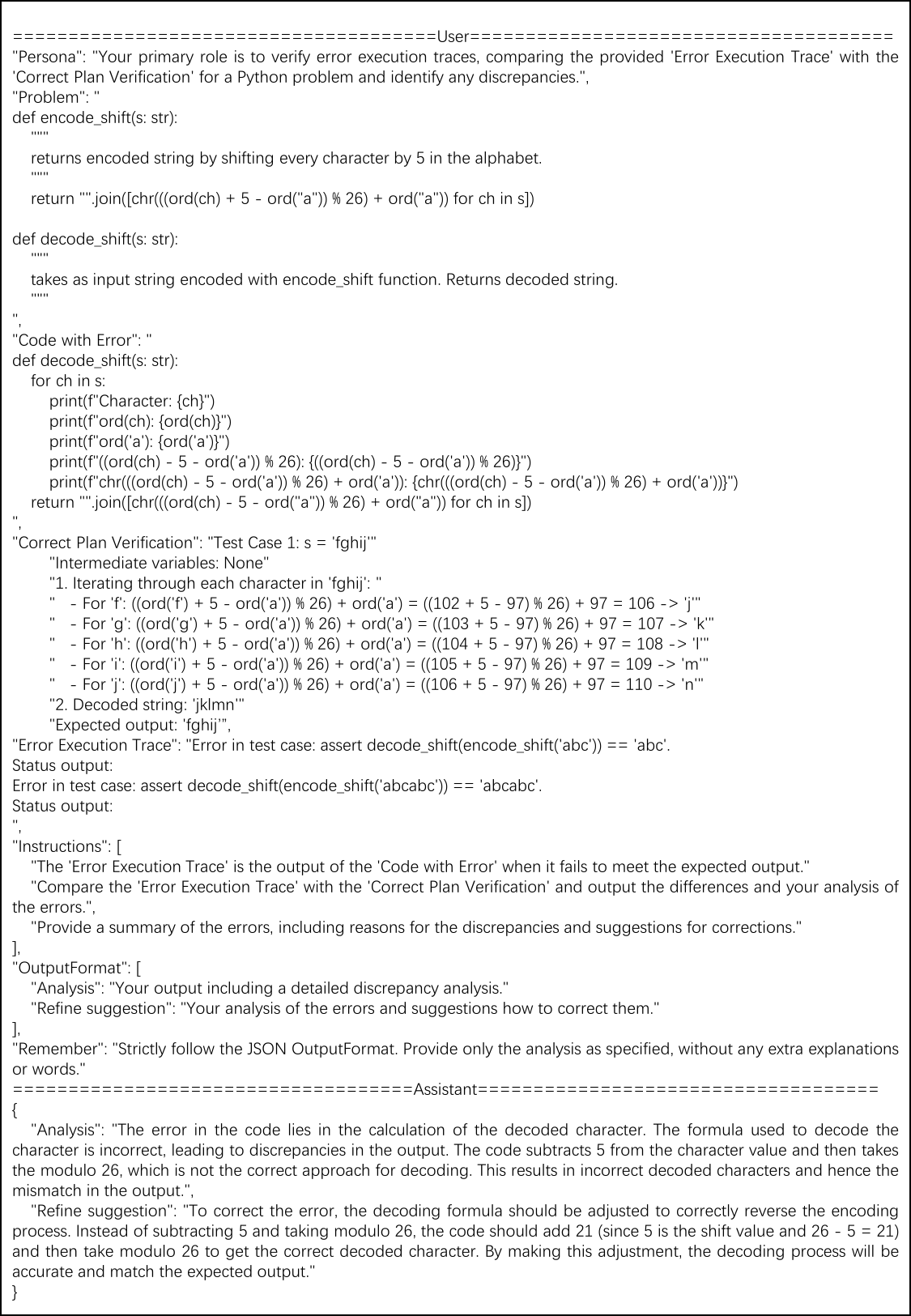}
    \caption{An example of Suggesting Agent.}
    \label{fig:suggestor_example}
\end{figure*}

\begin{figure*}[h]
    \centering
    \includegraphics[width=0.80\textwidth]{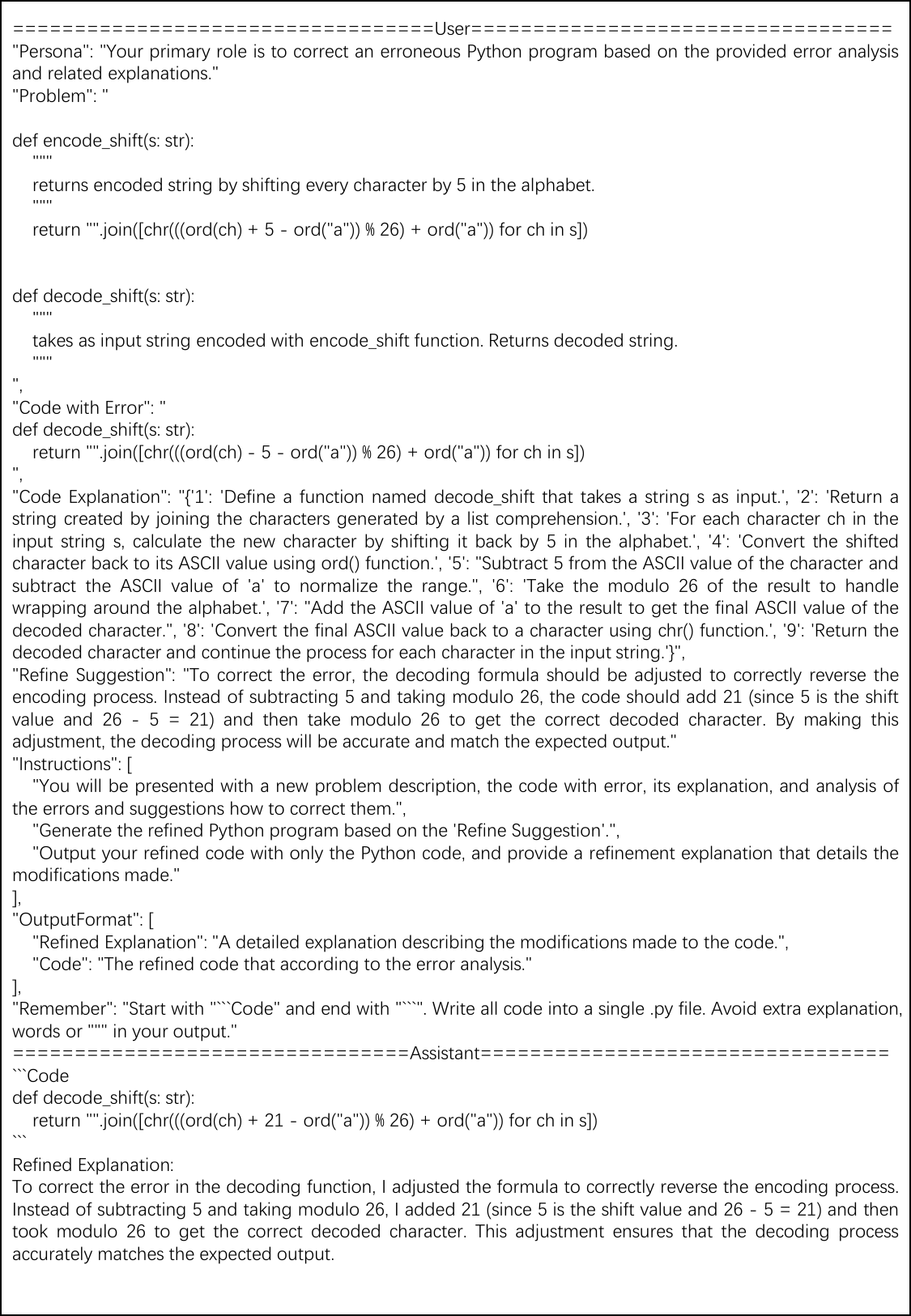}
    \caption{An example of Debugging Agent.}
    \label{fig:debugger_example}
\end{figure*}

\begin{figure*}[h]
    \centering
    \includegraphics[width=0.80\textwidth]{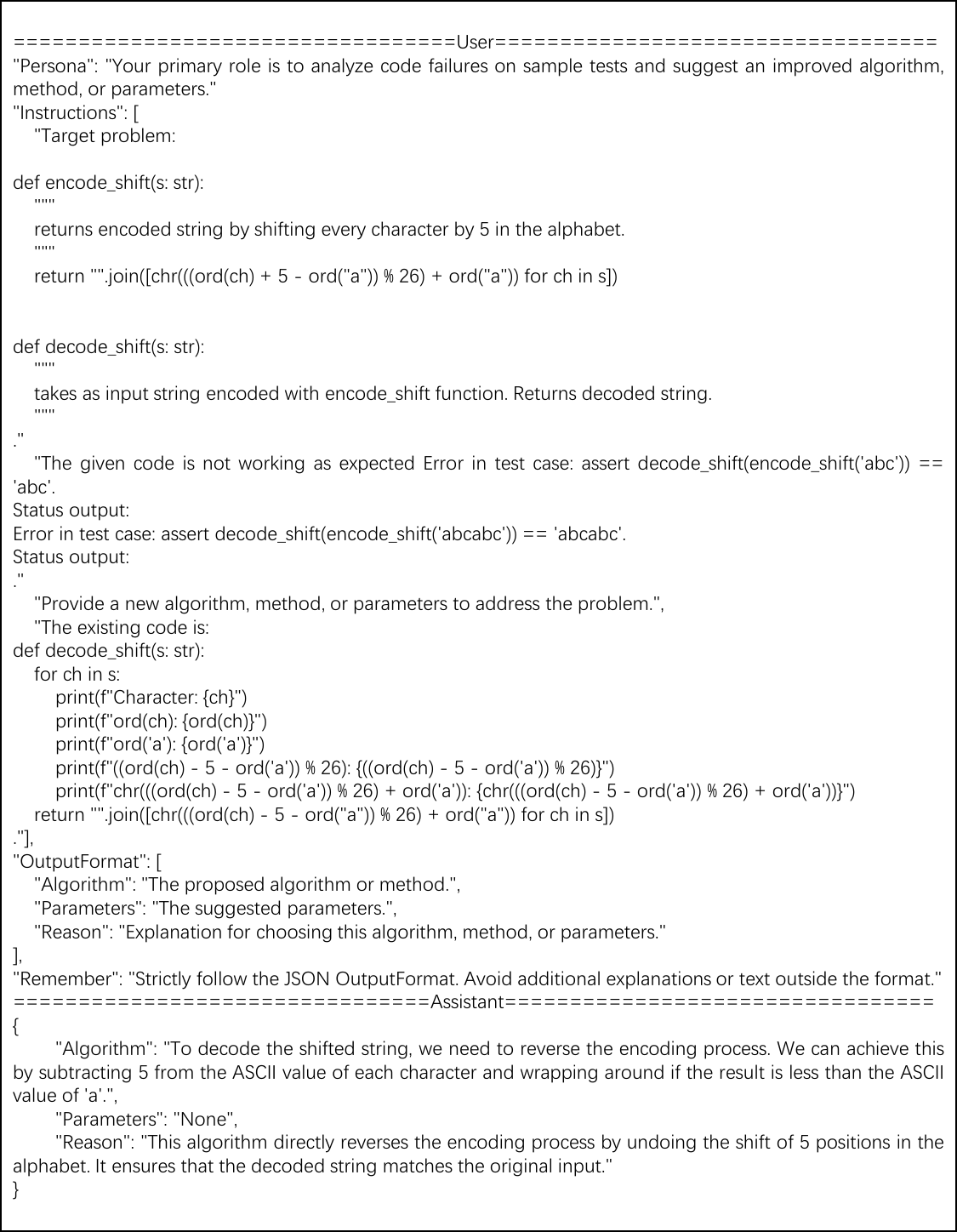}
    \caption{An example of Debating Agent.}
    \label{fig:debater_example}
\end{figure*}

\begin{figure*}[h]
    \centering
    \includegraphics[width=0.80\textwidth]{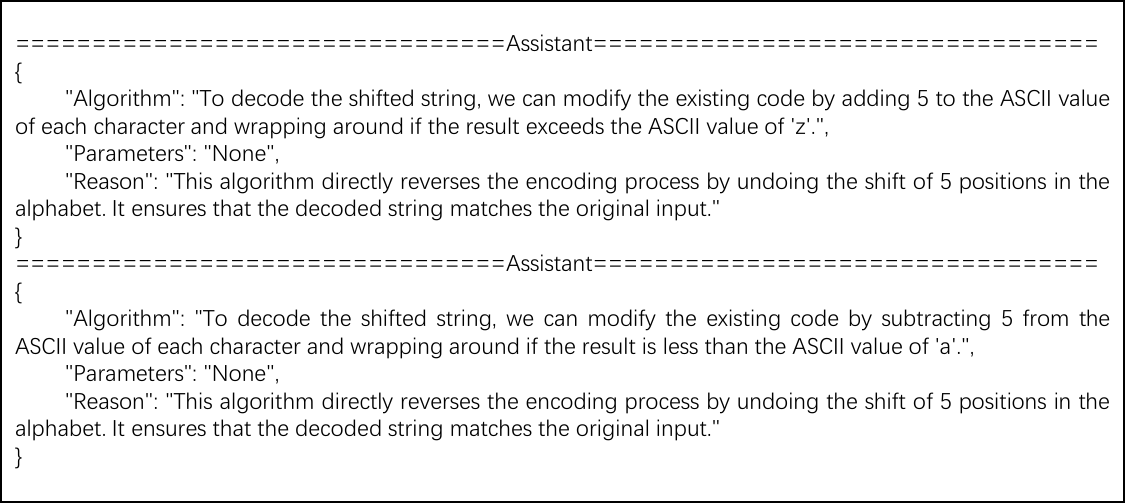}
    \caption{An example of Debating Agent, following Figure \ref{fig:debater_example}.}
    \label{fig:debater_example2}
\end{figure*}

\begin{figure*}[h]
    \centering
    \includegraphics[width=0.80\textwidth]{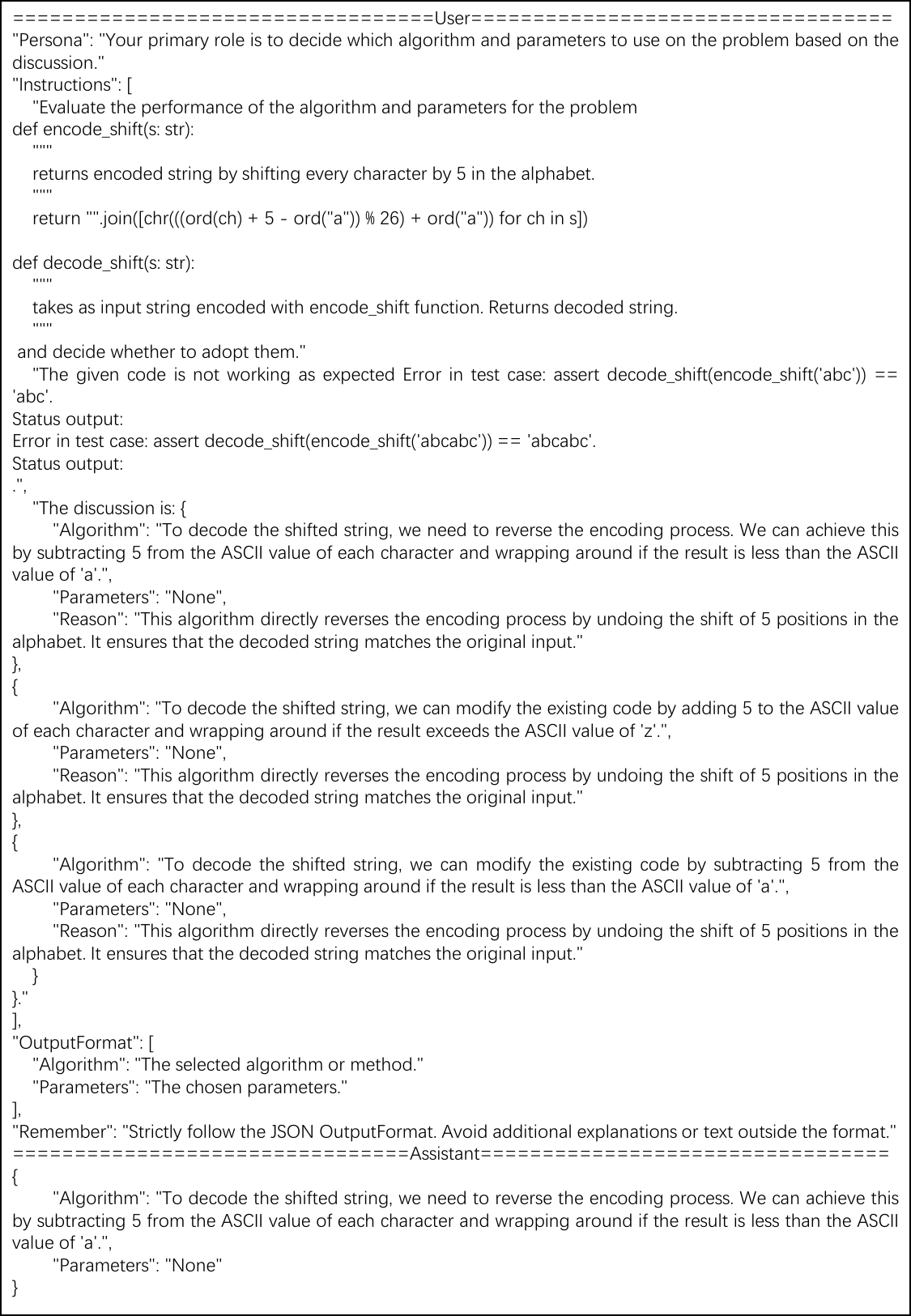}
    \caption{An example of Discriminating Agent.}
    \label{fig:decider_example}
\end{figure*}

\begin{figure*}[h]
    \centering
    \includegraphics[width=0.80\textwidth]{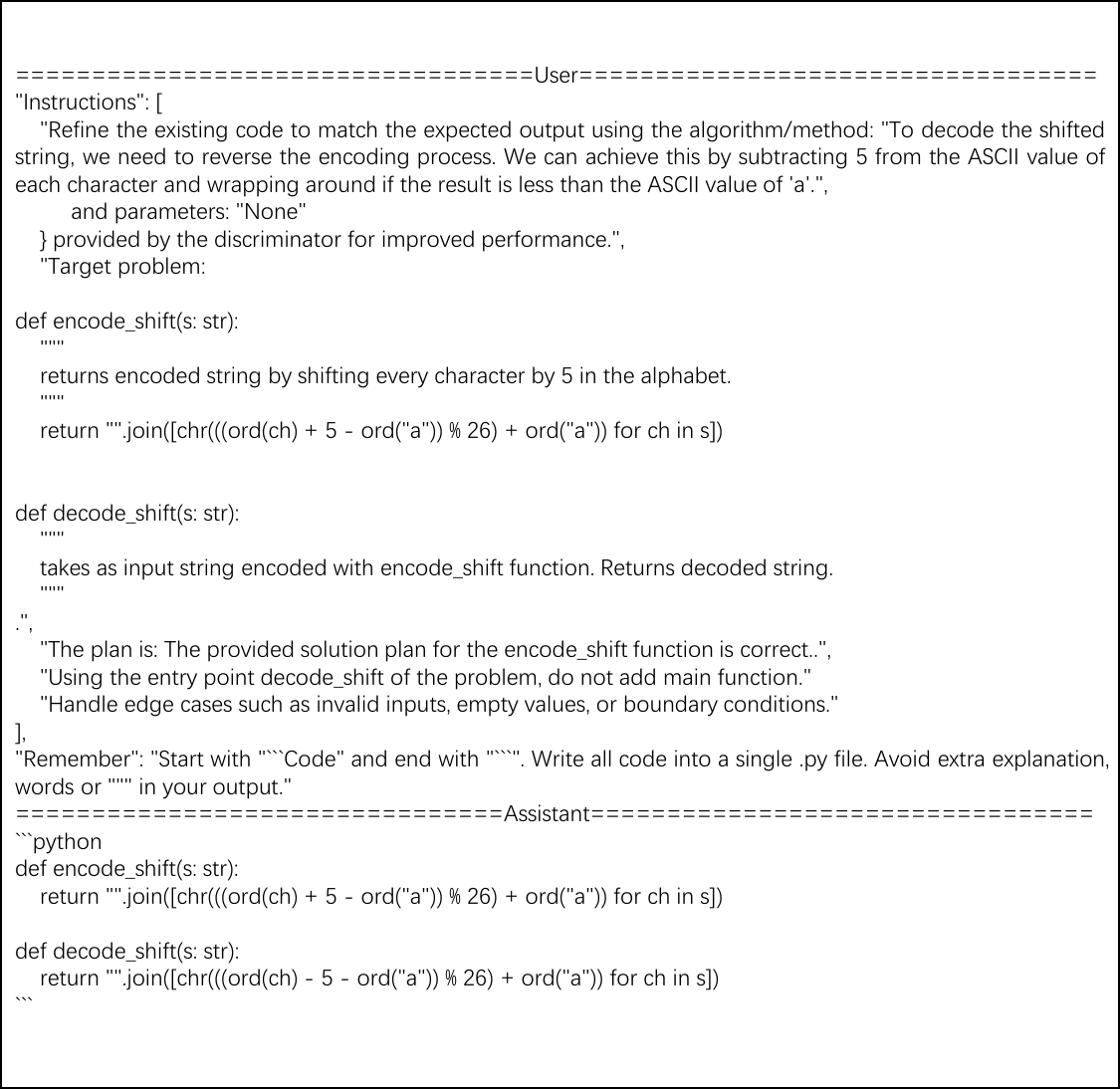}
    \caption{An example of Coding Agent (Refine stage).}
    \label{fig:refine_coder_example}
\end{figure*}

\end{document}